\documentclass[12pt]{article}
\usepackage{titling}
\usepackage[utf8]{inputenc}
\usepackage{amsmath}
\usepackage{amsthm}
\usepackage{mathtools}
\usepackage{amssymb}
\usepackage{mathrsfs}
\usepackage[english]{babel}
\usepackage{changepage}
\usepackage{graphicx}
\usepackage{mathrsfs}
\usepackage{relsize}
\usepackage{cite}
\usepackage{authblk}
\usepackage{appendix}
\usepackage{longtable}
\usepackage{accents} 
\graphicspath{ {./images/} }
\usepackage{fullpage}
\allowdisplaybreaks

\usepackage[normalem]{ulem}
\usepackage[dvipsnames]{xcolor}

\begin{document}

\title{Evolutionary games on networks with negative edge-based environmental feedback}
\author{\normalsize Katherine Betz$^1$,\; Feng Fu$^{2,3}$, \;and\;Naoki Masuda$^{4,5,6}$}
\date{%
\footnotesize$^1$ Department of Mathematics, State University of New York at Buffalo, NY 14260-2900, USA\\[1ex]
\footnotesize$^2$ Department of Mathematics, Dartmouth College, Hanover, NH 03755, USA\\[1ex]
\footnotesize$^3$ Department of Biomedical Data Science, Geisel School of Medicine at Dartmouth, Lebanon, NH 03755, USA\\[1ex]
\footnotesize$^4$ Gilberg S.\,Omenn Department of Computational Medicine and Bioinformatics, University of Michigan, MI 48109, USA\\[1ex]
\footnotesize$^5$ Department of Mathematics, University of Michigan, MI 48109, USA\\[1ex]
\footnotesize$^6$ Center for Computational Social Science, Kobe University, Kobe 657-8501, Japan
}
\maketitle

\begin{abstract}
Environmental feedback can fundamentally alter the game payoff structure underlying the evolution of cooperation. How feedback-evolving games interact with population structure under non-weak selection remains incompletely understood. We study an evolutionary game on networks in which each edge has a binary environmental state that determines the payoff matrix used by the two connected individuals. By assumption, cooperation tends to improve the local environment, which in turn favors defection, thereby creating negative feedback between strategies coupled with the environmental condition. We analyze the model under strong selection using a mean-field approximation, a pair approximation that tracks both strategy pairs and edge environments, and agent-based simulations. In well-mixed populations, the feedback can generate bistability between full cooperation and full defection, as well as sustained oscillations arising through a Hopf bifurcation. Sparse networks (i.e., with low nodes' degrees) can eliminate bistability, suppress oscillations, and produce stable intermediate levels of cooperation. The pair approximation captures these effects and generally agrees more closely with simulations on sparse networks than the mean-field approximation. Our results show that edge-based environmental feedback and network structure, such as density of edges, interact to determine whether the fraction of cooperators stabilizes or oscillates over time.
\end{abstract}

\textbf{keywords}: evolutionary game theory, feedback-evolving games, networks, oscillations, bifurcation, pair approximation

\section{Introduction}

%\kb{evolutionary game theory and mechanisms}
How cooperation emerges and persists among self-interested individuals is a central question in evolutionary biology and society at large. Evolutionary game theory provides a framework for studying how strategies spread when payoffs depend on interactions among individuals \cite{Hofbauer98, Nowak06B, Sigmund10}. In a well-mixed population, natural selection favors defection over cooperation in prototypical social dilemmas such as the prisoner's dilemma and public goods games. Cooperation in these settings therefore requires an additional mechanism that improves its prospects, such as kin selection, direct or indirect reciprocity, group selection, or network structure \cite{Nowak06S}. In this paper, we focus on feedback between strategies and the environment and on network structure, particularly on how these two mechanisms interact.

%\kb{feedback-evolving games}
In a feedback-evolving game, the payoff matrix changes over time in response to individuals' actions. Such feedback arises in ecological metacommunities, where species traits alter habitat conditions that shape competition and dispersal \cite{Leibold18, Liao22}; in insect populations, where foraging and colonization change resource and habitat availability \cite{Hanski11, Stella22HB}; and in microbial populations, where secreted public goods and metabolic byproducts alter the chemical environment and the benefits available to different strains \cite{West06, Sanchez13, Estrela18, Cordero12, Menon15, Lewin-Epstein17}. Examples in human societies include shared grasslands, where grazing decisions alter resource conditions that determine the success of grazing strategies \cite{Levin13}; the coevolution of social behavior and dispersal, which shapes kin structure in humans and other social organisms \cite{Mullon18}; and decision-making strategies that alter the environment determining their own success \cite{Rand17}. A seminal study coupled replicator dynamics to a slowly changing environmental resource influenced by individuals' recent actions, producing sustained oscillations between cooperative and uncooperative states~\cite{Weitz16}. Subsequent studies have incorporated autonomous resource renewal or decay and generalized the framework to arbitrary two-strategy games with environment-dependent payoffs \cite{Tilman20, Wang20}. Other extensions consider finite \cite{Lin19} or growing \cite{Bairagya21} populations; imitation, aspiration, or replicator--mutator dynamics \cite{Arefin21, Gong22, Huang25}; multiplayer and public goods games \cite{Shao19, Wang20, Jiang23}; and asymmetric and bimatrix games \cite{Hauert19, Shu22, Qin20, Liu20}. Further feedback mechanisms include payoff-dependent rates of environmental enhancement and degradation and fluctuations in local and global environments \cite{Cao21, Yan21, Ma24, Su19GT, Jiang23}. A growing body of work also uses control, incentives, and reinforcement learning to steer these systems toward cooperation \cite{Gong22, Han24, Yang25,
Ding23, Zhang23}.

%\kb{spatial and networked feedback}
Network structure can promote cooperation by allowing cooperators to form clusters that protect them from exploitation by defectors, a mechanism known as network reciprocity \cite{Nowak92, Ohtsuki06N, Szabo07, Allen17}. Environmental feedback can reinforce this mechanism, although their combined effects depend on how the environment and evolutionary dynamics are modeled \cite{Stella22PD, Szolnoki18, Wu18EP, Yang21, Lin19, Wu19, Lu23, Huang25, Guo23}. Existing spatial and network models commonly assign the environment either to the population as a whole or to individual nodes or neighborhoods. In the former case, a single environmental variable is shared by all players, while interactions occur on a network \cite{Gong18, Kawano19, Liu20, Shu22, Stella22PD, Zhang23, Wang25}. The benefits of population structure are then conditional: cooperation can be favored below a threshold degree, with oscillations arising at higher degrees \cite{Stella22PD}, and whether larger or smaller neighborhoods better maintain a replete environment depends on the payoff structure and the rates of environmental enhancement and degradation \cite{Wang25}. In the latter case, each individual or neighborhood has its own environment, determined by local cooperation or resource availability \cite{Jin18, Szolnoki18, Wu18EP, Lin19, Hauert19, Wu19, Yang21, Lu23}. Such models often report enhanced cooperation across different network structures \cite{Szolnoki18, Wu19, Lu23}, although this outcome is not universal \cite{Yang21}. The interaction between spatial structure and environmental feedback can also produce propagating cooperative wave fronts \cite{Lin19}, while the effect of connectivity on cooperation can depend on the strategy-update rule \cite{Huang25}. Comparisons of local and global feedback further show that network heterogeneity and mean degree can alter strategy--environment dynamics \cite{Zhong26}. These studies demonstrate the importance of coupling population structure with environmental feedback, but leave open how these effects change when environments are assigned to individual interactions.

%\kb{open questions and our model.}
In earlier work, we extended the two-community setting of Refs.~\cite{Gong18, Kawano19, Shu22} by allowing interactions within and between communities and assigning continuous environmental variables to each edge type \cite{Betz24}. This model produced multistability and anti-synchronous oscillations between communities. It is a coarse-grained counterpart of fully edge-based models, in which each edge has its own environmental variable \cite{Su19GT, He23, Zhu23}. However, the effects of individual-edge feedback on evolutionary dynamics beyond two communities remain unclear. Existing edge-based studies either derive weak-selection conditions for cooperation \cite{Su19GT} or simulate evolutionary dynamics on square lattices \cite{He23, Zhu23}; none analyzes oscillations and multistability generated by negative feedback, key phenomena in feedback-evolving games \cite{Weitz16}. Here, we extend our two-community model \cite{Betz24} in two directions. First, we consider a network of $N$ nodes with arbitrary topology, negative feedback, finite selection strength, and a tunable rate of environmental updating relative to strategy updating. Second, we replace the continuous environment with a binary state: each edge is either depleted, where mutual cooperation is the unique Nash equilibrium, or rich, where mutual defection is the unique Nash equilibrium. This simplification enables a pair approximation that tracks the joint states of adjacent nodes \cite{Matsuda92, Keeling99, Baalen00, Ohtsuki06JTB, Ohtsuki06N, KissMillerSimon17} together with each edge's environmental state. Using mean-field and pair approximations and agent-based simulations, we examine how local interactions in structured populations change the dynamics relative to well-mixed populations.

\section{Model}\label{sec:model}

Consider an evolutionary game on a static undirected unweighted network with $N$ nodes.  Each player, represented by a node, chooses either of the two actions, i.e., cooperation or defection, which can be represented as vectors 
$s=\begin{pmatrix}
1&0
\end{pmatrix}^\top$ and $s=\begin{pmatrix}
0&1
\end{pmatrix}^\top$, respectively, where ${}^{\top}$ represents the transposition.  Consider two game environments for pairwise interactions specified by the corresponding payoff matrices given by
$A_0 = \begin{pmatrix}R_0&S_0\\ T_0&P_0\end{pmatrix}$ and 
$A_1 = \begin{pmatrix}R_1&S_1\\ T_1&P_1\end{pmatrix}$.
Each edge $e$ is endowed with a binary environmental state that determines which of the two payoff matrices is used on $e$. We define edge's environmental state as $m \in \{ 0, 1 \}$. 

We assume that $A_0$ is the game played in the depleted environment, i.e., $m=0$, and that cooperation is the unique Nash equilibrium of this game, i.e., $R_0>T_0$ and $S_0>P_0$, which gives the harmony game. On the other hand, $A_1$ is the game played in a rich environment, i.e., $m=1$. We assume that defection is the unique Nash equilibrium of this latter game, i.e., $R_1<T_1$ and $S_1<P_1$, which gives the prisoner's dilemma game. We label these inequalities as 
\begin{align}
R_0>T_0,\;S_0>P_0,\;&\;R_1<T_1,\;S_1<P_1.
\label{inequal}
\end{align}

We denote by $\Omega_i$ the set of neighbors of the $i$th node. We define $\Omega_{i0}\subseteq\Omega_i$ and $\Omega_{i1}\subseteq\Omega_i$ to be the set of neighbors of the $i$th node with which the $i$th node plays game 0 (i.e., $m = 0$) and game 1 (i.e., $m = 1$), respectively.  It should be noted that $\Omega_{i0}\cup\Omega_{i1}=\Omega_i$ and $\Omega_{i0}\cap\Omega_{i1} = \emptyset$. Using these notations, we define the average payoff for node $i$ (i.e., the average over all neighbors of $i$), denoted by $F_i$, as 
\begin{equation}\label{payoffi}
F_i = \frac{1}{|\Omega_i|}\left[\sum_{j \in \Omega_{i0}} s_i^{\top} A_0 s_j + \sum_{j \in \Omega_{i1}} s_i^{\top} A_1 s_j\right].
\end{equation}

\begin{figure}[t!]
\centering 
\includegraphics[width=170mm]{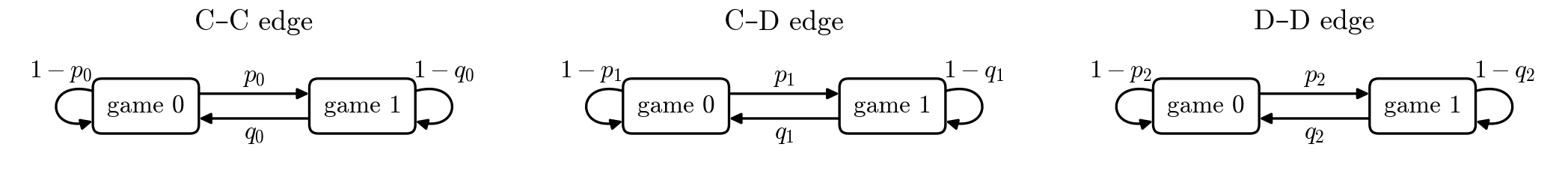} 
\caption{Schematic of the edge's environmental state updating.} 
\label{fig:schematic}
\end{figure}

Each round of the evolutionary dynamics consists of the following steps. We assume that $0 < \omega < 1$. With probability $1-\omega$, we carry out a strategy-updating step under the Fermi rule \cite{Blume93, Szabo98, Nowak04, Traulsen06PVE}. In one step of the strategy updating,
we select a node $i$ with strategy $s$ uniformly at random and then another node $j$ with strategy $s^\prime$ out of $i$'s neighbors, i.e., $\Omega_i$, uniformly at random. Then, $i$ adopts $j$'s strategy with probability
\begin{equation}\label{fermi}
\mathbb{P}(s\to s^\prime)=\frac{1}{1+e^{-\beta(F_{s^\prime}-F_s)}},
\end{equation}
where $\beta\geq 0$ represents the intensity of selection.

With probability $\omega$, a node $i$ is selected uniformly at random and all edges incident to $i$ independently update their environmental state.  If $m=0$ for such an edge, then $m$ switches to $m=1$ with probability $p_j$, where $j=0$ if $e$ is a cooperator-cooperator (CC) edge, $j=1$ if $e$ is a cooperator-defector (CD) edge, and $j=2$ if $e$ is a defector-defector (DD) edge. 
See Fig.~\ref{fig:schematic} for a schematic.
Then, with probability $1-p_j$, $m$ does not switch such that $m=0$.  Likewise, 
if $m=1$, then $m$ switches to $m=0$ with probability $q_j$. With probability $1-q_j$, $m$ does not switch such that $m=1$. We assume that 
\begin{equation}\label{probineq}
1 > p_0>p_1>p_2 > 0 \quad \text{and} \quad 0 < q_0<q_1<q_2 < 1.
\end{equation}

\section{Mean-field approximation} \label{sec:MF}

\subsection{Dynamics}

We assume that the population is well mixed (i.e., complete graph) in this section. Let us define $x$ as the fraction of cooperators and $1-x$ as the fraction of defectors in the population. We denote by $P_{CC}^{(m)}$, $P_{CD}^{(m)}$, and $P_{DD}^{(m)}$ the fractions of physical undirected edges of type $CC$, $CD$, and $DD$, respectively, that have environmental state $m\in\{0,1\}$. Therefore,
\begin{equation}\label{pnorm}
\sum_{m\in\{0,1\}}\left(P_{CC}^{(m)}+P_{CD}^{(m)}+P_{DD}^{(m)}\right)=1,
\end{equation}
and
\begin{equation}\label{xnorm}
x=\sum_{m\in\{0,1\}}\left(P_{CC}^{(m)}+\frac{1}{2}P_{CD}^{(m)}\right).
\end{equation}

We use the mean-field assumption that every node of each type (i.e., C or D) has the same proportion of edges of each type. Precisely speaking, all cooperators have the same number of C neighbors and that of D neighbors. Among their C neighbors, all cooperators are assumed to have the same number of CC edges in game 0. Then, for example, all cooperators have the same number of CC edges in game 1.

Let $M_{XY}^{(m)}$ be the fraction of $XY$ edges that have environmental state $m$, where $X, Y \in \{ \mathrm{C}, \mathrm{D} \}$. On the complete graph, the total fractions of the three strategy-pair types are
\begin{align}
\sum_{m\in\{0,1\}}P_{CC}^{(m)}&=\frac{x(Nx-1)}{N-1},\label{eq:totalCCfinite}\\
\sum_{m\in\{0,1\}}P_{CD}^{(m)}&=\frac{2Nx(1-x)}{N-1},\label{eq:totalCDfinite}\\
\sum_{m\in\{0,1\}}P_{DD}^{(m)}&=\frac{(1-x)[N(1-x)-1]}{N-1}.\label{eq:totalDDfinite}
\end{align}
Consequently, we obtain
\begin{align}
M_{CC}^{(m)}&=\frac{(N-1)P_{CC}^{(m)}}{x(Nx-1)},\label{conCC}\\
M_{CD}^{(m)}&=\frac{(N-1)P_{CD}^{(m)}}{2Nx(1-x)},\label{conCD}\\
M_{DD}^{(m)}&=\frac{(N-1)P_{DD}^{(m)}}{(1-x)[N(1-x)-1]}.\label{conDD}
\end{align}
We note that
\begin{equation}
\sum_{m\in\{0,1\}}M_{XY}^{(m)}=1,\quad
XY \in \{ CC, CD, DD \}.
\label{eq:MCC-MCD-MDD}
\end{equation}

The mean-field payoffs of a cooperator and a defector are
\begin{align}
F_C&=\frac{(Nx-1)R_0M_{CC}^{(0)}+N(1-x)S_0M_{CD}^{(0)}+(Nx-1)R_1M_{CC}^{(1)}+N(1-x)S_1M_{CD}^{(1)}}{N-1}
\end{align}
and
\begin{align}
F_D&=\frac{NxT_0M_{CD}^{(0)}+[N(1-x)-1]P_0M_{DD}^{(0)}+NxT_1M_{CD}^{(1)}+[N(1-x)-1]P_1M_{DD}^{(1)}}{N-1},
\end{align}
respectively. Equivalently, using Eq.~\eqref{eq:MCC-MCD-MDD}, we obtain
\begin{align}
F_C&=\frac{(Nx-1)\left[(R_0-R_1)M_{CC}^{(0)}+R_1\right]+N(1-x)\left[(S_0-S_1)M_{CD}^{(0)}+S_1\right]}{N-1},\label{eq:FCfinite2}\\
F_D&=\frac{Nx\left[(T_0-T_1)M_{CD}^{(0)}+T_1\right]+[N(1-x)-1]\left[(P_0-P_1)M_{DD}^{(0)}+P_1\right]}{N-1}.\label{eq:FDfinite2}
\end{align}

Conditioned on a strategy-update event, the probability that a defector samples a cooperator and adopts cooperation is
\begin{equation}\label{eq:Tplusfinite}
T^+=\frac{N}{N-1}x(1-x)\frac{1}{1+\exp[-\beta(F_C-F_D)]},
\end{equation}
and the probability that a cooperator samples a defector and adopts defection is
\begin{equation}\label{eq:Tminusfinite}
T^-=\frac{N}{N-1}x(1-x)\frac{1}{1+\exp[-\beta(F_D-F_C)]}.
\end{equation}
The factor $N/(N-1)$ accounts for sampling a neighbor without allowing a node to sample itself. Because $\Delta x=1/N$ and one update event corresponds to $\Delta t=1/N$, we obtain
\begin{equation}
\dot{x}\approx(1-\omega)\frac{N}{N-1}x(1-x)\tanh\left[\frac{\beta}{2}(F_C-F_D)\right].
\end{equation}
Therefore, in the large-population limit, we obtain
\begin{equation}\label{eq:dx/dt}
\dot{x}=(1-\omega)x(1-x)\tanh\left[\frac{\beta}{2}(F_C-F_D)\right].
\end{equation}
For weak selection, $0<\beta\ll1$, Eq.~\eqref{eq:dx/dt} becomes the replicator equation given by
\begin{equation}\label{eq:dxweak}
\dot{x}\approx\frac{\beta}{2}(1-\omega)x(1-x)(F_C-F_D).
\end{equation}

In the remainder of this section, we consider the limit of $N\to\infty$. Equations~\eqref{conCC}--\eqref{conDD} become 
\begin{align}
M_{CC}^{(m)}&=\frac{P_{CC}^{(m)}}{x^2},\label{MCCi}\\
M_{CD}^{(m)}&=\frac{P_{CD}^{(m)}}{2x(1-x)},\label{MCDi}\\
M_{DD}^{(m)}&=\frac{P_{DD}^{(m)}}{(1-x)^2}.\label{MDDi}
\end{align}
For $0<x<1$, the payoffs can therefore be written as
\begin{align}
F_C&=\frac{(R_0-R_1)P_{CC}^{(0)}}{x}+R_1x+\frac{(S_0-S_1)P_{CD}^{(0)}}{2x}+S_1(1-x),\label{eq:FCinf}\\
F_D&=\frac{(T_0-T_1)P_{CD}^{(0)}}{2(1-x)}+T_1x+\frac{(P_0-P_1)P_{DD}^{(0)}}{1-x}+P_1(1-x).\label{eq:FDinf}
\end{align}

Let $L_{XY}^{(m)}=|E|P_{XY}^{(m)}$ be the number of undirected $XY$ edges in environmental state $m$, where $|E|$ is the number of edges in the network. Conditioned on an environmental-update event, a physical edge is selected when either of its two endpoints is selected, which occurs with probability $2/N$. Therefore, to leading order in $N$, we obtain
\begin{align}
\mathbb{E}\!\left[\Delta L_{CC}^{(0)}\mid\mathrm{env}\right]
&=\frac{2}{N}\left[-p_0L_{CC}^{(0)}+q_0L_{CC}^{(1)}\right]\notag\\
&\approx(N-1)\left[-p_0P_{CC}^{(0)}+q_0\left(x^2-P_{CC}^{(0)}\right)\right],\label{eq:envCC}\\
\mathbb{E}\!\left[\Delta L_{CD}^{(0)}\mid\mathrm{env}\right]
&\approx(N-1)\left\{-p_1P_{CD}^{(0)}+q_1\left[2x(1-x)-P_{CD}^{(0)}\right]\right\},\label{eq:envCD}\\
\mathbb{E}\!\left[\Delta L_{DD}^{(0)}\mid\mathrm{env}\right]
&\approx(N-1)\left\{-p_2P_{DD}^{(0)}+q_2\left[(1-x)^2-P_{DD}^{(0)}\right]\right\},\label{eq:envDD}
\end{align}
where ``env'' means the environmental-update event, and $\approx$ means ``approximately equal to.''

We define
\begin{equation}\label{eq:rhoMF}
\rho=\frac{1}{1+\exp[-\beta(F_C-F_D)]},
\end{equation}
which is the mean-field probability of the transition $D\to C$ across a sampled CD edge. The probability of the reverse transition
$C \to D$ is equal to $1-\rho$.
Conditioned on a strategy-update event, let $T^+=x(1-x)\rho$ and $T^-=x(1-x)(1-\rho)$. To leading order in $N$, the expected changes in the edge counts are
\begin{align}
\mathbb{E}\!\left[\Delta L_{CC}^{(m)}\mid\mathrm{strat}\right]
&\approx(N-1)\left[\frac{T^+P_{CD}^{(m)}}{2(1-x)}-\frac{T^-P_{CC}^{(m)}}{x}\right],\label{eq:stratCC}\\
\mathbb{E}\!\left[\Delta L_{CD}^{(m)}\mid\mathrm{strat}\right]
&\approx(N-1)\left\{\frac{T^+P_{DD}^{(m)}}{1-x}-\frac{[xT^++(1-x)T^-]P_{CD}^{(m)}}{2x(1-x)}+\frac{T^-P_{CC}^{(m)}}{x}\right\},\label{eq:stratCD}\\
\mathbb{E}\!\left[\Delta L_{DD}^{(m)}\mid\mathrm{strat}\right]
&\approx(N-1)\left[\frac{T^-P_{CD}^{(m)}}{2x}-\frac{T^+P_{DD}^{(m)}}{1-x}\right],
\label{eq:stratDD}
\end{align}
where ``strat'' means the strategy-update event.
Because $|E|=N(N-1)/2$ and $\Delta t=1/N$, Eqs.~\eqref{eq:envCC}--\eqref{eq:stratDD} yield the following four-dimensional mean-field dynamical system:
\begin{align}
\dot{x}&=(1-\omega)x(1-x)(2\rho-1),\label{eq:dxrho}\\
\dot{P}_{CC}^{(0)}
&=2\omega \left[ q_0x^2-(p_0+q_0)P_{CC}^{(0)}\right] \notag\\
&\quad +(1-\omega)\left[xP_{CD}^{(0)}\rho-2(1-x)P_{CC}^{(0)}(1-\rho)\right],\label{eq:dpcc}\\
\dot{P}_{CD}^{(0)}
&=2\omega\left[ 2q_1x(1-x)-(p_1+q_1)P_{CD}^{(0)}\right] \notag\\
&\quad +(1-\omega)\left\{x\left[2P_{DD}^{(0)}-P_{CD}^{(0)}\right]\rho+(1-x)\left[2P_{CC}^{(0)}-P_{CD}^{(0)}\right](1-\rho)\right\},\label{eq:dpcd}\\
\dot{P}_{DD}^{(0)}
&=2\omega \left[q_2(1-x)^2-(p_2+q_2)P_{DD}^{(0)}\right] \notag\\
&\quad +(1-\omega)\left[ (1-x)P_{CD}^{(0)}(1-\rho)-2xP_{DD}^{(0)}\rho\right].\label{eq:dpdd}
\end{align}
Equation~\eqref{eq:dxrho} is identical to Eq.~\eqref{eq:dx/dt}. The feasible region is given by
$0\leq P_{CC}^{(0)}\leq x^2$, $0\leq P_{CD}^{(0)}\leq2x(1-x)$, and $0\leq P_{DD}^{(0)}\leq(1-x)^2$.
We obtain the densities of edges in the environmental state $1$ from the corresponding total pair densities, e.g., $P_{CC}^{(1)}=x^2-P_{CC}^{(0)}$.

\subsection{Equilibria}\label{sec:equilibria}

We analyze the equilibria of the four-dimensional mean-field dynamics given by Eqs.~\eqref{eq:dxrho}, \eqref{eq:dpcc}, \eqref{eq:dpcd}, and \eqref{eq:dpdd}. The Jacobian in the interior of the feasible region is given in Appendix~\ref{app:MF1}. Because the payoff expressions in Eqs.~\eqref{eq:FCinf} and \eqref{eq:FDinf} contain ratios that become indeterminate at $x=0$ or $x=1$, we treat boundary stability using the regularized conditional environmental fractions introduced below. This gives an approximate rare-invader calculation.

If $x^\ast=0$, then Eq.~\eqref{xnorm} implies $P_{CC}^{(0)\ast}=P_{CD}^{(0)\ast}=0$. Equation~\eqref{eq:dpdd} then gives
\begin{equation}\label{zeroeq}
\left(x^\ast,P_{CC}^{(0)\ast},P_{CD}^{(0)\ast},P_{DD}^{(0)\ast}\right)
=\left(0,0,0,\frac{q_2}{p_2+q_2}\right).
\end{equation}
Similarly, if $x^\ast=1$, then $P_{CD}^{(0)\ast}=P_{DD}^{(0)\ast}=0$, and Eq.~\eqref{eq:dpcc} gives
\begin{equation}\label{oneeq}
\left(x^\ast,P_{CC}^{(0)\ast},P_{CD}^{(0)\ast},P_{DD}^{(0)\ast}\right)
=\left(1,\frac{q_0}{p_0+q_0},0,0\right).
\end{equation}

For $0<x<1$, we write
\begin{align}
P_{CC}^{(0)}&=c_1x^2,\label{eq:PC0subs}\\
P_{CD}^{(0)}&=2c_2x(1-x),\label{eq:PB0subs}\\
P_{DD}^{(0)}&=c_3(1-x)^2.\label{eq:PD0subs}
\end{align}
Then, we obtain $c_1=M_{CC}^{(0)}$, $c_2=M_{CD}^{(0)}$, and $c_3=M_{DD}^{(0)}$, and each $c_j$ lies between $0$ and $1$. In terms of these variables, we write
\begin{align}
F_C&=x\left[(R_0-R_1)c_1+R_1\right]+(1-x)\left[(S_0-S_1)c_2+S_1\right],\label{eq:FCreg}\\
F_D&=x\left[(T_0-T_1)c_2+T_1\right]+(1-x)\left[(P_0-P_1)c_3+P_1\right],\label{eq:FDreg}
\end{align}
and
\begin{equation}\label{eq:rhoreg}
\rho=\frac{1}{1+\exp[-\beta(F_C-F_D)]}.
\end{equation}
Substituting Eqs.~\eqref{eq:PC0subs}--\eqref{eq:PD0subs} into the mean-field equations and applying the product rule gives the regularized mean-field dynamical system: 
\begin{align}
\dot{x}&=(1-\omega)x(1-x)(2\rho-1),\label{eq:regx}\\
\dot{c}_1&=2\omega\left[q_0-(p_0+q_0)c_1\right]
+2(1-\omega)(1-x)\rho(c_2-c_1),\label{eq:regc1}\\
\dot{c}_2&=2\omega\left[q_1-(p_1+q_1)c_2\right]+(1-\omega)\left[x(1-\rho)(c_1-c_2)+(1-x)\rho(c_3-c_2)\right],\label{eq:regc2}\\
\dot{c}_3&=2\omega\left[q_2-(p_2+q_2)c_3\right]
+2(1-\omega)x(1-\rho)(c_2-c_3).\label{eq:regc3}
\end{align}
Unlike the original coordinates, Eqs.~\eqref{eq:regx}--\eqref{eq:regc3} have finite limits at $x=0$ and $x=1$. The variables $c_j$ associated with pair types that are absent at a boundary are auxiliary rare-invader quantities. We estimate them by steady states of the corresponding boundary equations. This is the approximation used in the following invasion analysis.

We denote by $c_{j,0}^\ast$ a steady solution of Eqs.~\eqref{eq:regc1}--\eqref{eq:regc3} at the all-defector equilibrium, satisfying $x^\ast = 0$. We obtain
\begin{align}
c_{1,0}^\ast&=\frac{\omega q_0+(1-\omega)\rho_0^\ast c_{2,0}^\ast}
{\omega(p_0+q_0)+(1-\omega)\rho_0^\ast},\label{eq:c10}\\
c_{2,0}^\ast&=\frac{2\omega q_1+(1-\omega)\rho_0^\ast c_{3,0}^\ast}
{2\omega(p_1+q_1)+(1-\omega)\rho_0^\ast},\label{eq:c20}\\
c_{3,0}^\ast&=\frac{q_2}{p_2+q_2},\label{eq:c30}
\end{align}
where
\begin{equation}\label{eq:rho0star}
\rho_0^\ast=\frac{1}{1+\exp\left\{-\beta\left[(S_0-S_1)c_{2,0}^\ast+S_1-(P_0-P_1)c_{3,0}^\ast-P_1\right]\right\}}.
\end{equation}
Equations~\eqref{eq:c20} and \eqref{eq:rho0star} form an implicit scalar fixed-point problem. If it has more than one solution, each solution defines a candidate rare-cooperator invasion calculation.

The Jacobian of the dynamical system given by Eqs.~\eqref{eq:regx}--\eqref{eq:regc3} at such a boundary solution is triangular after a permutation of the auxiliary variables. Its four eigenvalues are
\begin{align}
\lambda_{\mathrm{inv}}^{(0)}&=(1-\omega)(2\rho_0^\ast-1),\label{eq:lambda0inv}\\
\lambda_{c_1}^{(0)}&=-2\omega(p_0+q_0)-2(1-\omega)\rho_0^\ast,\label{eq:lambda0c1}\\
\lambda_{c_2}^{(0)}&=-2\omega(p_1+q_1)+(1-\omega)\left[-\rho_0^\ast+\beta(S_0-S_1)\rho_0^\ast(1-\rho_0^\ast)(c_{3,0}^\ast-c_{2,0}^\ast)\right],\label{eq:lambda0c2}\\
\lambda_{c_3}^{(0)}&=-2\omega(p_2+q_2).\label{eq:lambda0c3}
\end{align}
Here, $\lambda_{\mathrm{inv}}^{(0)}$ is the leading growth rate of a small cooperator fraction. The eigenvalues shown in Eqs.~\eqref{eq:lambda0c1} and \eqref{eq:lambda0c3} are negative. Therefore, within this approximation, the all-defector equilibrium is locally attracting when
\begin{equation}\label{x=0express}
\lambda_{\mathrm{inv}}^{(0)}<0
\qquad\text{and}\qquad
\lambda_{c_2}^{(0)}<0.
\end{equation}
By substituting Eq.~\eqref{eq:lambda0inv} in $\lambda_{\mathrm{inv}}^{(0)}<0$, we obtain
$\rho_0^* < \tfrac{1}{2}$.
Using Eq.~\eqref{eq:rho0star}, we find that the condition $\rho_0^* < \tfrac{1}{2}$ is equivalent to $F_C^{(0)} < F_D^{(0)}$, that is, a rare
cooperator cannot invade when it is less fit than the resident defectors.

We denote by $c_{j,1}^\ast$ a steady solution of Eqs.~\eqref{eq:regc1}--\eqref{eq:regc3} at the all-cooperator equilibrium, satisfying $x^\ast =1$. We obtain
\begin{align}
c_{1,1}^\ast&=\frac{q_0}{p_0+q_0},\label{eq:c11}\\
c_{2,1}^\ast&=\frac{2\omega q_1+(1-\omega)(1-\rho_1^\ast)c_{1,1}^\ast}
{2\omega(p_1+q_1)+(1-\omega)(1-\rho_1^\ast)},\label{eq:c21}\\
c_{3,1}^\ast&=\frac{\omega q_2+(1-\omega)(1-\rho_1^\ast)c_{2,1}^\ast}
{\omega(p_2+q_2)+(1-\omega)(1-\rho_1^\ast)},\label{eq:c31}
\end{align}
where
\begin{equation}\label{eq:rho1star}
\rho_1^\ast=\frac{1}{1+\exp\left\{-\beta\left[(R_0-R_1)c_{1,1}^\ast+R_1-(T_0-T_1)c_{2,1}^\ast-T_1\right]\right\}}.
\end{equation}
Again, Eqs.~\eqref{eq:c21} and \eqref{eq:rho1star} define an implicit scalar fixed-point problem. The eigenvalues of the regularized boundary Jacobian are
\begin{align}
\lambda_{\mathrm{inv}}^{(1)}&=(1-\omega)(1-2\rho_1^\ast),\label{eq:lambda1inv}\\
\lambda_{c_1}^{(1)}&=-2\omega(p_0+q_0),\label{eq:lambda1c1}\\
\lambda_{c_2}^{(1)}&=-2\omega(p_1+q_1)+(1-\omega)\left[-(1-\rho_1^\ast)+\beta(T_0-T_1)\rho_1^\ast(1-\rho_1^\ast)(c_{1,1}^\ast-c_{2,1}^\ast)\right],\label{eq:lambda1c2}\\
\lambda_{c_3}^{(1)}&=-2\omega(p_2+q_2)-2(1-\omega)(1-\rho_1^\ast).\label{eq:lambda1c3}
\end{align}
The quantity $\lambda_{\mathrm{inv}}^{(1)}$ is the growth rate of a small defector fraction. The eigenvalues shown in Eqs.~\eqref{eq:lambda1c1} and \eqref{eq:lambda1c3} are negative. Therefore, the all-cooperator equilibrium is locally attracting under the same approximation when
\begin{equation}\label{x=1express}
\lambda_{\mathrm{inv}}^{(1)}<0
\qquad\text{and}\qquad
\lambda_{c_2}^{(1)}<0.
\end{equation}
By substituting Eq.~\eqref{eq:lambda1inv} in $\lambda_{\mathrm{inv}}^{(1)}<0$, we obtain
$\rho_1^* > \tfrac{1}{2}$.
Using Eq.~\eqref{eq:rho1star}, we find that the condition $\rho_1^* > \tfrac{1}{2}$ is equivalent to $F_C^{(0)} > F_D^{(0)}$, that is, a rare
defector cannot invade when it is less fit than the resident cooperators.

\begin{figure}
\centering
\includegraphics[width=150mm]{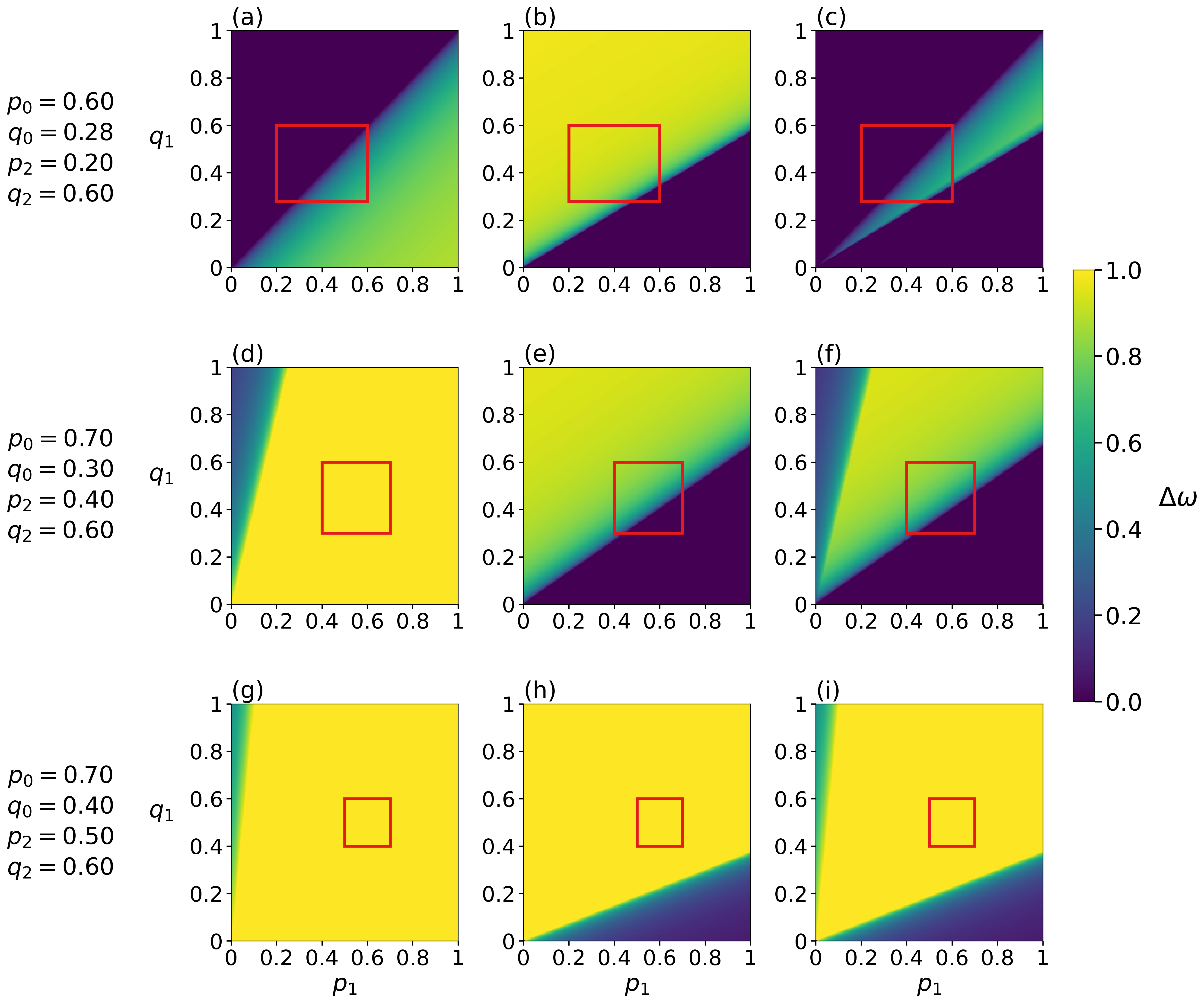}
\caption{Ranges of $\omega$ for which the boundary equilibria are stable as functions of $p_1$ and $q_1$. 
For panels (a)--(c), we set $p_0=0.6$, $q_0=0.28$, $p_2=0.6$, and $q_2=0.6$. For panels (d)--(f), we set $p_0=0.7$, $q_0=0.3$, $p_2=0.4$, and $q_2=0.6$. For panels (g)--(i), we set $p_0=0.7$, $q_0=0.4$, $p_2=0.5$, and $q_2=0.6$. The first column (i.e., panels (a), (d), and (g)) shows $\Delta \omega$, which is the range of $\omega \in [0, 1]$ values (equivalently, the fraction of $\omega$ values) for which the all-defector equilibrium is stable. The second column (i.e., panels (b), (e), and (h)) shows the regions in which the all-cooperator equilibrium is stable. The third column (i.e., panels (c), (f), and (i)) shows the size of the intersection of the two $\omega$ regions in the first two columns, i.e., the fraction of $\omega$ values for which all-defector and all-cooperator equilibria are bistable. We set $\beta=5$. The red rectangle in each panel marks the admissible region satisfying the probability constraints given by Eq.~\eqref{probineq}.}
\label{fig:bistability}
\end{figure}

We first explore parameter regions in which the all-defector equilibrium or the all-cooperator equilibrium is stable when we set some of the parameters to arbitrary fixed values. For this and all the following numerical simulations, we set
$\begin{pmatrix}R_0&S_0\\T_0&P_0\end{pmatrix}
=\begin{pmatrix}5&1\\3&0\end{pmatrix}$ and
$\begin{pmatrix}R_1&S_1\\T_1&P_1\end{pmatrix}
=\begin{pmatrix}3&0\\4&2\end{pmatrix}$, which satisfy Eq.~\eqref{inequal}. 
In Fig.~\ref{fig:bistability}(a), (d), and (g),
we show the ranges of $\omega$ for which both conditions in Eq.~\eqref{x=0express} are satisfied, i.e., when the all-defector equilibrium is stable, for three combinations of $p_0$, $q_0$, $p_2$, and $q_2$ values. We vary $p_1$ and $q_1$. The color scale
indicates the fraction of $\omega$ values for which the all-defector equilibrium is stable for given ($p_0$, $q_0$, $p_1$, $q_1$, $p_2$, $q_2$).
The red rectangle indicates the admissible region, i.e., the ($p_1$, $q_1$) region that satisfies Eq.~\eqref{probineq}. We find that 
the all-defector equilibrium is stable for some combinations of $p_1$, $q_1$, and $\omega$ for the $p_0$, $q_0$, $p_2$, and $q_2$ values used in 
Fig.~\ref{fig:bistability}(a) and all combinations of feasible $p_1$ and $q_1$ values and $\omega \in (0, 1)$ in 
Fig.~\ref{fig:bistability}(d) and (g).
Similarly, Fig.~\ref{fig:bistability}(b), (e), and (h) shows the ranges of $\omega$ for which both conditions in Eq.~\eqref{x=1express} are satisfied, i.e., when the all-cooperator equilibrium is stable. The overlap of these parameter regions, shown in Fig.~\ref{fig:bistability}(c), (f), and (i), identifies the parameter regions in which both all-defector and all-cooperator equilibria are stable (i.e., satisfying both Eqs.~\eqref{x=0express} and \eqref{x=1express}), thus showing bistability under the mean-field approximation. We find that the bistability is common in this model.

For an interior equilibrium with $0<x^\ast<1$, Eq.~\eqref{eq:regx} requires
$\rho^\ast=1/2$, which is equivalent to $F_C^\ast=F_D^\ast$.
At $\rho^\ast=1/2$, the three remaining equilibrium equations are linear in $c_1^\ast$, $c_2^\ast$, and $c_3^\ast$:
\begin{equation}\label{eq:interiorlinear}
\begin{pmatrix}
2\omega(p_0+q_0)+(1-\omega)(1-x^\ast ) & -(1-\omega)(1-x^\ast) & 0\\
-\dfrac{1-\omega}{2}x^\ast & 2\omega(p_1+q_1)+\dfrac{1-\omega}{2} & -\dfrac{1-\omega}{2}(1-x^\ast)\\
0 & -(1-\omega)x^\ast & 2\omega(p_2+q_2)+(1-\omega)x^\ast
\end{pmatrix}
\begin{pmatrix}c_1^\ast \\c_2^\ast \\c_3^\ast \end{pmatrix}
=2\omega\begin{pmatrix}q_0\\q_1\\q_2\end{pmatrix}.
\end{equation}
Because $0<\omega<1$ and $p_j+q_j>0$, $j\in \{ 0, 1, 2 \}$, by assumption, the coefficient matrix in Eq.~\eqref{eq:interiorlinear} is strictly diagonally dominant and therefore nonsingular. Therefore, for a trial value of $x^\ast \in(0,1)$, we first solve Eq.~\eqref{eq:interiorlinear} and then impose 
$\rho^\ast= 1/2$. This reduces the interior-equilibrium calculation to a scalar root-finding problem in $x^\ast$. Only roots for which $0\leq c_j^\ast\leq1$ for all $j$ are admissible. We then evaluate the Jacobian in Appendix~\ref{app:MF1} at each resulting equilibrium.

\begin{figure}[t]
\centering
\includegraphics[width=130mm]{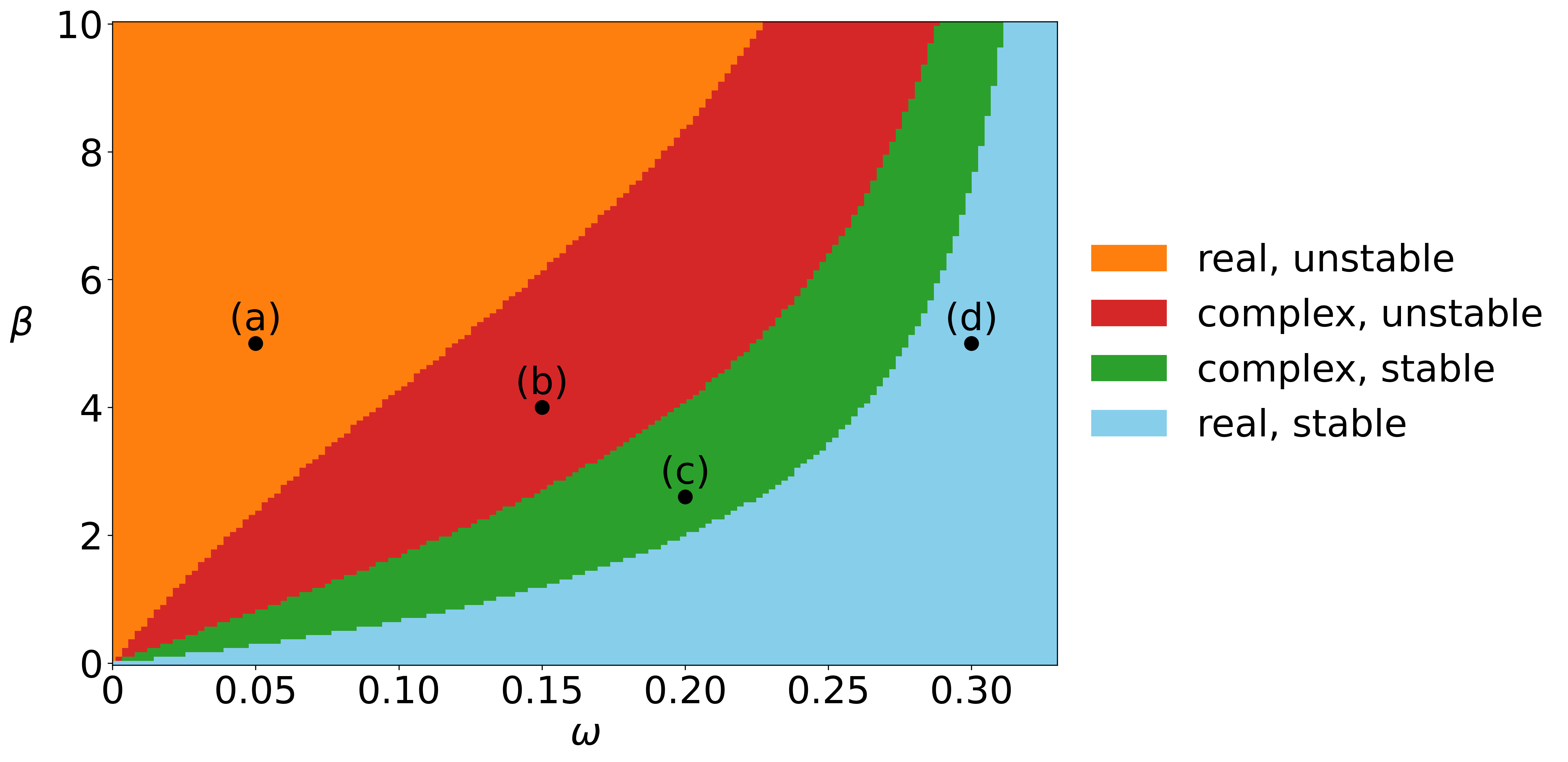}
\caption{Classification of the leading eigenvalues of the interior equilibrium when $\omega$ and $\beta$ are varied. We set $p_0=0.8$, $q_0=0.3$, $p_1=0.4$, $q_1=0.5$, $p_2=0.3$, and $q_2=0.7$ in this and all the following figures. The orange region corresponds to real eigenvalues with positive real parts (unstable). The red region corresponds to complex-conjugate eigenvalues with positive real parts (spiral source). The green region corresponds to complex-conjugate eigenvalues with negative real parts (spiral sink). The blue region corresponds to real eigenvalues with negative real parts (stable). The labeled points indicate the parameter values used in Figs.~\ref{fig:osc(a)comparisonchart}--\ref{fig:osc(d)comparisonchart}: (a) $(\omega,\beta)=(0.05,5)$, (b) $(\omega,\beta)=(0.15,4)$, (c) $(\omega,\beta)=(0.2,2.6)$, and (d) $(\omega,\beta)=(0.3,5)$.}
\label{fig:phasedia}
\end{figure}
 
To examine the stability and robustness the internal equilibrium as parameters vary, we set $p_0=0.8$, $q_0=0.3$, $p_1=0.4$, $q_1=0.5$, $p_2=0.3$, and $q_2=0.7$, which altogether satisfy Eq.~\eqref{probineq}, and vary $\omega$ and $\beta$. We use the same values of $p_j$ and $q_j$, $j\in \{ 0, 1, 2 \}$, in the numerical simulations of the remainder of this manuscript as well. For this parameter set, the system has two interior equilibria. We focus on the one with the larger $x^\ast$ value because we have verified that the lower equilibrium is a saddle over the parameter range considered. Figure~\ref{fig:phasedia} classifies the leading eigenvalues of the retained, upper equilibrium in the $(\omega,\beta)$ plane. The figure reveals four regimes, marked by the representative points (a)--(d). For small $\omega$, shown in orange and represented by point (a), the upper internal equilibrium is unstable with a leading eigenvalue with positive real part. As $\omega$ increases, the leading eigenvalues become a complex-conjugate pair with positive real part, shown in red and represented by point (b), so trajectories near the equilibrium spiral outward. The boundary between the unstable and stable complex-eigenvalue regions (i.e., between the red and green regions) is where this eigenvalue pair crosses the imaginary axis, which marks a Hopf bifurcation. When $\omega$ is larger than the Hopf bifurcation curve, the equilibrium is a stable focus, resulting in damped oscillations (shown in green and represented by point (c)). For still larger $\omega$, the eigenvalues become real, and the equilibrium is a stable node (shown in blue and represented by point (d)). All three boundaries shift to larger $\omega$ as $\beta$ increases. Therefore, under stronger selection, edge environments must update more frequently to stabilize the interior equilibrium.

\section{Pair approximation}\label{sec:pa}

To derive the pair approximation, we consider a static undirected $k$-regular graph, for which $|E|=Nk/2$, where we remind that $|E|$ is the number of edges in the network. Throughout this section, $P_{XY}^{(m)}$, where $XY\in\{CC,CD,DD\}$ and $m\in\{0,1\}$, denotes the fraction of physical undirected edges with endpoint strategies $XY$ and environmental state $m$.  Thus, we obtain
\begin{equation}\label{pa:def-P}
P_{XY}^{(m)}=\frac{L_{XY}^{(m)}}{|E|},
\end{equation}
where we remind that $L_{XY}^{(m)}$ is the number of undirected edges corresponding to $P_{XY}^{(m)}$. This is the same edge-density convention as in section~\ref{sec:MF}; in particular, a $CD$ edge is counted once rather than once per orientation.  Therefore, we use Eqs.~\eqref{pnorm} and \eqref{xnorm} as in section~\ref{sec:MF}. Although Eq.~\eqref{pa:def-P} uses undirected edges, a strategy update samples an ordered focal--neighbor node pair. Let $\Pi_{XY}^{(m)}$ denote the probability that the focal node has strategy $X$, its sampled neighbor has strategy $Y$, and their edge has environmental state $m$. Counting oriented stubs gives
\begin{equation}\label{pa:ordered-pairs}
\Pi_{CC}^{(m)}=P_{CC}^{(m)},\qquad
\Pi_{CD}^{(m)}=\Pi_{DC}^{(m)}=\frac{1}{2}P_{CD}^{(m)},\qquad
\Pi_{DD}^{(m)}=P_{DD}^{(m)}.
\end{equation}
Equation~\eqref{pa:ordered-pairs} is the conversion that prevents factors of two from being mixed between ordered update events and undirected pair densities.

\subsection{Conditional probabilities and pair closure}\label{pa:closure-subsection}

For $X,Y\in\{C,D\}$, let $q_{Y\mid X}^{(m)}$ be the probability that an edge emanating from a focal $X$ node directly connects to a $Y$ node and has environmental state $m$.  Stub counting yields
\begin{align}
q_{C\mid C}^{(m)}&=\frac{P_{CC}^{(m)}}{x},
\label{qcccon}\\
q_{D\mid C}^{(m)}&=\frac{P_{CD}^{(m)}}{2x},\\
q_{C\mid D}^{(m)}&=\frac{P_{CD}^{(m)}}{2(1-x)},\\
q_{D\mid D}^{(m)}&=\frac{P_{DD}^{(m)}}{1-x},
\label{qddcon}
\end{align}
where we remind that $x$ and $1-x$ are the fraction of cooperators and defectors, respectively, in the population.
Thus, we obtain
\begin{equation}\label{pa:q-normalization}
\sum_{m \in \{ 0, 1 \}} \left(q_{C\mid X}^{(m)}+q_{D\mid X}^{(m)}\right)=1,
\qquad X\in\{C,D\},
\end{equation}
and
\begin{equation}\label{pa:mixed-sampling}
(1-x)q_{C\mid D}^{(m)}=x q_{D\mid C}^{(m)}=\frac{1}{2}P_{CD}^{(m)}.
\end{equation}
A mixed edge (i.e., an edge joining one $C$ node and one $D$ node) can be sampled in either direction; Eq.~\eqref{pa:mixed-sampling} shows that each orientation has probability $P_{CD}^{(m)}/2$.

We use the standard pair closure \cite{Matsuda92, Keeling99, Baalen00, Ohtsuki06JTB, KissMillerSimon17}. That is, after conditioning on the strategy at the center of a path, we treat the states of distinct edges attached to the same node as independent. In conditional-probability notation, we write
\begin{equation}
\label{pa:wedge-closure}
\mathbb{P}\!\left((A,m_1),(C,m_2)\mid B\right)
\approx q_{A\mid B}^{(m_1)}q_{C\mid B}^{(m_2)}.
\end{equation}
Repeated application of Eq.~\eqref{pa:wedge-closure} gives the multinomial neighborhood distribution below.

Consider a sampled mixed edge with environmental state $n\in\{0,1\}$.  Excluding this focal edge, we let
\begin{equation}\label{pa:kappa-vector}
\boldsymbol{\kappa}=\left(
\kappa_{CC}^{(0)},\kappa_{CC}^{(1)},
\kappa_{CD}^{(0)},\kappa_{CD}^{(1)},
\kappa_{DC}^{(0)},\kappa_{DC}^{(1)},
\kappa_{DD}^{(0)},\kappa_{DD}^{(1)}
\right),
\end{equation}
where each $\kappa_{XY}^{(m)}$ is an edge count.  For example, among the $k-1$ nonfocal edges connected to the $C$ endpoint, $\kappa_{CC}^{(0)}$ counts those whose other endpoint is $C$ and whose environmental state is $0$.  The first four entries count the remaining edges connected to the $C$ endpoint, and the last four entries count those connected to the $D$ endpoint.  The admissible count sets are
\begin{align}
\mathcal{K}_C&=\left\{\boldsymbol{\kappa}_C
=\left(
\kappa_{CC}^{(0)},\kappa_{CC}^{(1)},
\kappa_{CD}^{(0)},\kappa_{CD}^{(1)}
\right) :
\sum_{m \in \{ 0, 1 \}} \left(\kappa_{CC}^{(m)}+\kappa_{CD}^{(m)}\right)=k-1\right\},
\label{pa:KC}\\
\mathcal{K}_D&=\left\{\boldsymbol{\kappa}_D
=\left(
\kappa_{DC}^{(0)},\kappa_{DC}^{(1)},
\kappa_{DD}^{(0)},\kappa_{DD}^{(1)}
\right) :
\sum_{m \in \{ 0, 1 \}} \left(\kappa_{DC}^{(m)}+\kappa_{DD}^{(m)}\right)=k-1\right\}.
\label{pa:KD}
\end{align}
Under Eq.~\eqref{pa:wedge-closure}, the joint probability of the two endpoint neighborhoods is
\begin{align}
\mathcal{M}(\boldsymbol{\kappa})={}&
\frac{(k-1)!}{\displaystyle\prod_{m=0}^{1}
\kappa_{CC}^{(m)}!\,\kappa_{CD}^{(m)}!}
\prod_{m=0}^{1}
\left(q_{C\mid C}^{(m)}\right)^{\kappa_{CC}^{(m)}}
\left(q_{D\mid C}^{(m)}\right)^{\kappa_{CD}^{(m)}}
\notag\\
&\times
\frac{(k-1)!}{\displaystyle\prod_{m=0}^{1}
\kappa_{DC}^{(m)}!\,\kappa_{DD}^{(m)}!}
\prod_{m=0}^{1}
\left(q_{C\mid D}^{(m)}\right)^{\kappa_{DC}^{(m)}}
\left(q_{D\mid D}^{(m)}\right)^{\kappa_{DD}^{(m)}}.
\label{pa:multinomial}
\end{align}
For any function $g$ of the neighborhood counts, we use the shorthand
\begin{equation}\label{pa:expectation}
\left\langle g(\boldsymbol{\kappa})\right\rangle
=
\sum_{\boldsymbol{\kappa}_C\in\mathcal{K}_C}
\sum_{\boldsymbol{\kappa}_D\in\mathcal{K}_D}
\mathcal{M}(\boldsymbol{\kappa})g(\boldsymbol{\kappa}).
\end{equation}

\subsection{Payoffs and strategy-update probabilities}\label{sec:Ferm}

For a mixed focal edge in environmental state $n$, the average payoffs of its $C$ and $D$ endpoints are, respectively,
\begin{align}
F_C^{(n)}(\boldsymbol{\kappa})
&=\frac{1}{k}\left[S_n+\sum_{m \in \{ 0, 1 \}} \left(
\kappa_{CC}^{(m)}R_m+\kappa_{CD}^{(m)}S_m
\right)\right],
\label{pa:FC}\\
F_D^{(n)}(\boldsymbol{\kappa})
&=\frac{1}{k}\left[T_n+\sum_{m \in \{ 0, 1 \}} \left(
\kappa_{DC}^{(m)}T_m+\kappa_{DD}^{(m)}P_m
\right)\right].
\label{pa:FD}
\end{align}
The first term in each equation is the payoff obtained on the focal mixed edge. We define
\begin{align}
f_n^+(\boldsymbol{\kappa})
&=\frac{1}{1+\exp\!\left\{-\beta\left[F_C^{(n)}(\boldsymbol{\kappa})-F_D^{(n)}(\boldsymbol{\kappa})\right]\right\}},
\label{fermipa}\\
f_n^-(\boldsymbol{\kappa})
&=\frac{1}{1+\exp\!\left\{-\beta\left[F_D^{(n)}(\boldsymbol{\kappa})-F_C^{(n)}(\boldsymbol{\kappa})\right]\right\}}
=1-f_n^+(\boldsymbol{\kappa}).
\label{pa:fminus}
\end{align}
Here $f_n^+$ is the probability that the focal $D$ endpoint copies the $C$ endpoint, and $f_n^-$ is the probability of the reverse update.

The Fermi function is nonlinear.  Therefore, the pair approximation developed here retains the configuration averages $\langle f_n^\pm\rangle$ and the count-weighted averages $\langle\kappa_{XY}^{(m)}f_n^\pm\rangle$.  In general,
\begin{equation}\label{pa:nonlinear-warning}
\left\langle f_n^+(\boldsymbol{\kappa})\right\rangle
\neq
\frac{1}{1+\exp\{-\beta\langle F_C^{(n)}-F_D^{(n)}\rangle\}},
\end{equation}
and similarly $\langle\kappa f_n^\pm\rangle\neq\langle\kappa\rangle\langle f_n^\pm\rangle$.

\subsection{Contribution of an environmental-update event}

Each elementary update corresponds to $\Delta t=1/N$.  With probability $\omega$, an environmental update is carried out; with probability $1-\omega$, a strategy update is carried out.

Let
$r(CC)=0$, $r(CD)=1$, and $r(DD)=2$, corresponding to the definition of $p_j$ and $q_j$ (with $j \in \{0, 1, 2 \}$), which is schematically shown in Fig.~\ref{fig:schematic}.
A given undirected edge is updated when either of its two endpoint nodes is selected, which occurs with probability $2/N$. Therefore, conditioned on an environmental-update event, the expected change in $L_{XY}^{(m)}$, $m \in \{ 0, 1 \}$, induced by the environmental-update event is
\begin{align}
\mathbb{E}\!\left[\Delta L_{XY}^{(0)}\mid\mathrm{env}\right]
&=\frac{2}{N}\left[-p_{r(XY)}L_{XY}^{(0)}+q_{r(XY)}L_{XY}^{(1)}\right]
\notag\\
&=k\left[-p_{r(XY)}P_{XY}^{(0)}+q_{r(XY)}P_{XY}^{(1)}\right],
\label{pa:env-count-0}\\
\mathbb{E}\!\left[\Delta L_{XY}^{(1)}\mid\mathrm{env}\right]
&=-\mathbb{E}\!\left[\Delta L_{XY}^{(0)}\mid\mathrm{env}\right].
\label{pa:env-count-1}
\end{align}
The second equality in Eq.~\eqref{pa:env-count-0} follows from $2|E|/N=k$.  It also shows directly that the expected number of undirected edges updated in one environmental event is $k$, as required by the model rule.

\subsection{Contribution of a strategy-update event}

Conditioned on a strategy-update event, the probability of sampling a $D\to C$ ordered mixed edge in state $n$, observing configuration $\boldsymbol{\kappa}$, and completing the imitation is
\begin{equation}\label{pa:event-prob-plus}
\Pi_{DC}^{(n)}\mathcal{M}(\boldsymbol{\kappa})f_n^+(\boldsymbol{\kappa})
=\frac{P_{CD}^{(n)}}{2}\mathcal{M}(\boldsymbol{\kappa})f_n^+(\boldsymbol{\kappa}).
\end{equation}
The corresponding probability for a $C\to D$ event is
\begin{equation}\label{pa:event-prob-minus}
\Pi_{CD}^{(n)}\mathcal{M}(\boldsymbol{\kappa})f_n^-(\boldsymbol{\kappa})
=\frac{P_{CD}^{(n)}}{2}\mathcal{M}(\boldsymbol{\kappa})f_n^-(\boldsymbol{\kappa}).
\end{equation}
Let $\delta_{mn}$ denote the Kronecker delta.  For one completed $D\to C$ update, the changes in the numbers of undirected edges of state $m$ are
\begin{align}
\Delta L_{CC}^{(m)}&=\kappa_{DC}^{(m)}+\delta_{mn},
\label{pa:increments-plus}\\
\Delta L_{CD}^{(m)}&=\kappa_{DD}^{(m)}-\kappa_{DC}^{(m)}-\delta_{mn},\\
\Delta L_{DD}^{(m)}&=-\kappa_{DD}^{(m)}.
\end{align}
For one completed $C\to D$ update, they are
\begin{align}
\Delta L_{CC}^{(m)}&=-\kappa_{CC}^{(m)},\\
\Delta L_{CD}^{(m)}&=\kappa_{CC}^{(m)}-\kappa_{CD}^{(m)}-\delta_{mn},\\
\Delta L_{DD}^{(m)}&=\kappa_{CD}^{(m)}+\delta_{mn}.
\label{pa:increments-minus}
\end{align}
For example, when the focal $D$ becomes $C$, each of its nonfocal $D$--$C$ edges becomes a $C$--$C$ edge, each of its nonfocal $D$--$D$ edges becomes a $C$--$D$ edge, and the focal mixed edge becomes a $C$--$C$ edge.  Equations~\eqref{pa:increments-plus}--\eqref{pa:increments-minus} include that focal edge through $\delta_{mn}$.

Averaging the changes in the numbers of each edge type over the focal-edge state (i.e., $n$), the endpoint neighborhoods, and the imitation outcome provides the strategy-update drift.  We define $\mathcal{G}_{XY}^{(m)}$ to be the expected change in the number of undirected $XY$ edges in state $m$ in one strategy-update event. We obtain
\begin{align}
\mathcal{G}_{CC}^{(m)}
={}&\frac{1}{2}\sum_{n \in \{ 0, 1 \}} P_{CD}^{(n)}
\left\langle
\left(\kappa_{DC}^{(m)}+\delta_{mn}\right)f_n^+
-\kappa_{CC}^{(m)}f_n^-
\right\rangle,
\label{pa:GCC}\\
\mathcal{G}_{CD}^{(m)}
={}&\frac{1}{2}\sum_{n \in \{ 0, 1 \}} P_{CD}^{(n)}
\left\langle
\left(\kappa_{DD}^{(m)}-\kappa_{DC}^{(m)}-\delta_{mn}\right)f_n^+
% \right. \notag\\[-0.5ex]
% &\hspace{10em}\left.
+\left(\kappa_{CC}^{(m)}-\kappa_{CD}^{(m)}-\delta_{mn}\right)f_n^-
\right\rangle,
\label{pa:GCD}\\
\mathcal{G}_{DD}^{(m)}
={}&\frac{1}{2}\sum_{n \in \{ 0, 1 \}} P_{CD}^{(n)}
\left\langle
\left(\kappa_{CD}^{(m)}+\delta_{mn}\right)f_n^-
-\kappa_{DD}^{(m)}f_n^+
\right\rangle.
\label{pa:GDD}
\end{align}
The factor $1/2$ is the probability weight for each of the two orientations of an undirected $CD$ edge.  All $f_n^\pm$ in Eqs.~\eqref{pa:GCC}--\eqref{pa:GDD} are evaluated at $\boldsymbol{\kappa}$.  With these definitions, we obtain
\begin{equation}\label{pa:strategy-count-change}
\mathbb{E}\!\left[\Delta L_{XY}^{(m)}\mid\mathrm{strat}\right]
=\mathcal{G}_{XY}^{(m)},
\qquad XY\in\{CC,CD,DD\}.
\end{equation}

\subsection{Continuum limit and final equations}

Because $P_{XY}^{(m)}=L_{XY}^{(m)}/|E|$ and $\Delta t=1/N$, we obtain
\begin{align}
\dot P_{XY}^{(m)}
&=\lim_{N\to\infty}\frac{\mathbb{E}[\Delta P_{XY}^{(m)}]}{\Delta t}
\notag\\
&=\frac{N}{|E|}\left\{
\omega\mathbb{E}\!\left[\Delta L_{XY}^{(m)}\mid\mathrm{env}\right]
+(1-\omega)\mathbb{E}\!\left[\Delta L_{XY}^{(m)}\mid\mathrm{strat}\right]
\right\}
\notag\\
&=\frac{2}{k}\left\{
\omega\mathbb{E}\!\left[\Delta L_{XY}^{(m)}\mid\mathrm{env}\right]
+(1-\omega)\mathbb{E}\!\left[\Delta L_{XY}^{(m)}\mid\mathrm{strat}\right]
\right\}.
\label{pa:scaling}
\end{align}
Substitution of Eqs.~\eqref{pa:env-count-0}, \eqref{pa:env-count-1}, and \eqref{pa:strategy-count-change} into Eq.~\eqref{pa:scaling} gives the following six closed pair-approximation equations:
\begin{align}
\dot P_{CC}^{(0)}
&=2\omega\left[-p_0P_{CC}^{(0)}+q_0P_{CC}^{(1)}\right]
+\frac{2(1-\omega)}{k}\mathcal{G}_{CC}^{(0)},
\label{eq:dpcc0pa}\\
\dot P_{CC}^{(1)}
&=2\omega\left[p_0P_{CC}^{(0)}-q_0P_{CC}^{(1)}\right]
+\frac{2(1-\omega)}{k}\mathcal{G}_{CC}^{(1)},
\label{eq:dpcc1pa}\\
\dot P_{CD}^{(0)}
&=2\omega\left[-p_1P_{CD}^{(0)}+q_1P_{CD}^{(1)}\right]
+\frac{2(1-\omega)}{k}\mathcal{G}_{CD}^{(0)},
\label{eq:dpcd0pa}\\
\dot P_{CD}^{(1)}
&=2\omega\left[p_1P_{CD}^{(0)}-q_1P_{CD}^{(1)}\right]
+\frac{2(1-\omega)}{k}\mathcal{G}_{CD}^{(1)},
\label{eq:dpcd1pa}\\
\dot P_{DD}^{(0)}
&=2\omega\left[-p_2P_{DD}^{(0)}+q_2P_{DD}^{(1)}\right]
+\frac{2(1-\omega)}{k}\mathcal{G}_{DD}^{(0)},
\label{eq:dpdd0pa}\\
\dot P_{DD}^{(1)}
&=2\omega\left[p_2P_{DD}^{(0)}-q_2P_{DD}^{(1)}\right]
+\frac{2(1-\omega)}{k}\mathcal{G}_{DD}^{(1)}.
\label{pa:dpdd1}
\end{align}
Equations~\eqref{qcccon}--\eqref{qddcon}, \eqref{pa:KC}--\eqref{pa:fminus}, and \eqref{pa:GCC}--\eqref{pa:GDD} close these equations in terms of the pair densities.

The pair equations reproduce the node-density drift. Using Eq.~\eqref{xnorm}, we obtain
\begin{align}
\dot x
&=\sum_{m \in \{ 0, 1 \}} \left(\dot P_{CC}^{(m)}+\frac{1}{2}\dot P_{CD}^{(m)}\right)
\notag\\
&=\frac{1-\omega}{2}\sum_{n \in \{ 0, 1 \}} P_{CD}^{(n)}
\left\langle f_n^+-f_n^-\right\rangle
\notag\\
&=\frac{1-\omega}{2}\sum_{n \in \{ 0, 1 \}} P_{CD}^{(n)}
\left\langle 2f_n^+-1\right\rangle.
\label{pa:xdot}
\end{align}
Therefore, environmental switching does not change $x$, as expected.

Only five of the six pair densities are independent.  For example, one may integrate
$P_{CC}^{(0)}$, $P_{CC}^{(1)}$, $P_{CD}^{(0)}$, $P_{CD}^{(1)}$, $P_{DD}^{(0)}$,
and use
\begin{equation}\label{pa:eliminate}
P_{DD}^{(1)}
=1-P_{CC}^{(0)}-P_{CC}^{(1)}-P_{CD}^{(0)}-P_{CD}^{(1)}-P_{DD}^{(0)}.
\end{equation}
We then obtain the cooperator density $x$ from Eq.~\eqref{xnorm}.

\section{Agent-based simulations}

To test the accuracy of the mean-field theory and pair approximation, we run agent-based simulations of the evolutionary dynamics on the complete graph with $N=10^3$ nodes, a random $10$-regular graph with $N=10^4$ nodes, a random $4$-regular graph with $N=10^4$ nodes, and the square lattice with $N = 10^4$ nodes (with a linear size of $100$) and periodic boundary conditions. In each simulation run, we select a fraction $x(0)$ of the nodes uniformly at random and initialize them as cooperator. We initialize the remaining nodes as defector.

\begin{figure}[t!]
\centering
\includegraphics[width=140mm]{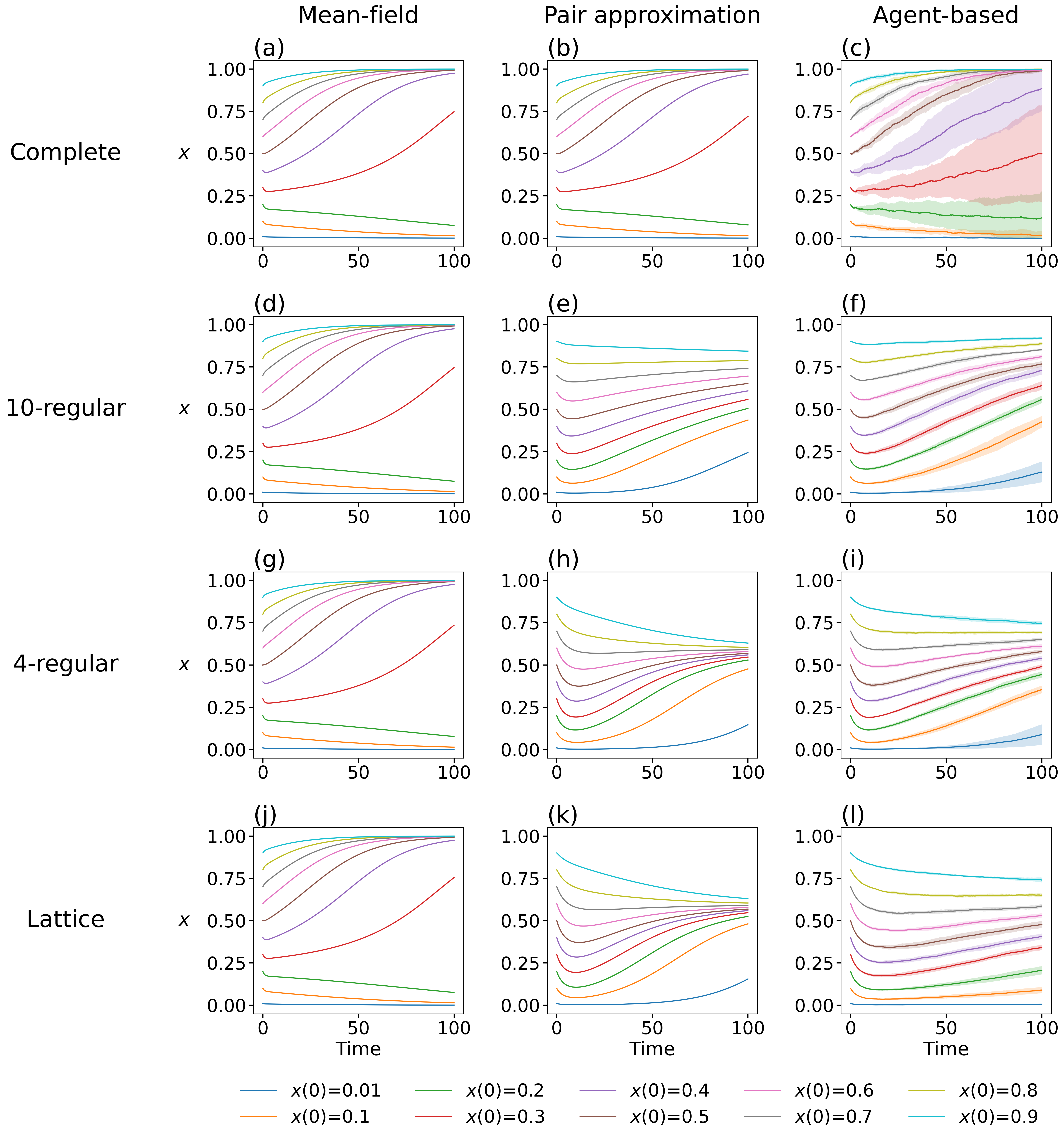}
\caption{Accuracy of the mean-field and pair approximation for monostable and bistable dynamics on four networks. 
(a)--(c): Complete graph of $10^3$ nodes. (d)--(f): A random $10$-regular graph of $10^4$ nodes. (g)--(i) A random $4$-regular graph of $10^4$ nodes. (j)--($\ell$): Square lattice of $10^4$ nodes. We use the same four networks in the following figures.
The left column (i.e., panels (a), (d), (g), and (j)) shows the mean-field approximation. The middle column (i.e., panels (b), (e), (h), and (k)) shows the pair approximation. The right column (i.e., panels (c), (f), (i), and (${\rm \ell}$)) shows the agent-based simulations; in these panels,
the lines show the average over ten runs, and the shaded regions indicate one standard deviation around the average.
We set $\omega = 0.7$, $\beta =5$, and $x(0) \in \{0.01, 0.1, 0.2, \ldots, 0.9 \}$.}
\label{fig:bicomparisonchart}
\end{figure}

We first examine how the dynamics depend on the initial condition. Figure~\ref{fig:bicomparisonchart} shows the fraction of cooperators $x(t)$ for $t \in [0,100]$ with initial fractions $x(0) \in \{0.01, 0.1, 0.2, \ldots, 0.9\}$, using $\omega = 0.7$ and $\beta = 5$. For the complete graph (Fig.~\ref{fig:bicomparisonchart}(a)--(c)), the mean-field and pair approximations agree closely with the agent-based simulations except for $x(0) = 0.3$. The mean-field approximation, pair approximation, and agent-based simulations all show bistability with this parameter set. In other words, trajectories starting from $x(0) \le 0.2$ approach full defection, whereas those starting from $x(0) \ge 0.4$ approach full cooperation. Near the threshold, at $x(0) = 0.3$, the agent-based simulations show large variability across runs because stochastic demographic fluctuations heavily influence which equilibrium is reached.

Because the mean-field approximation does not account for network structure, its trajectories are identical across all four networks (Fig.~\ref{fig:bicomparisonchart}(a), (d), (g), and (j)). The agent-based simulations on the random $10$-regular graph show that cooperation increases from every initial condition, including $x(0) = 0.01$, so the bistability is lost at least for this set of initial conditions (Fig.~\ref{fig:bicomparisonchart}(f)). In contrast, the mean-field approximation predicts bistability (Fig.~\ref{fig:bicomparisonchart}(d)). The pair approximation for the $10$-regular graph (Fig.~\ref{fig:bicomparisonchart}(e)) closely agrees with the agent-based simulations, including the apparent lack of bistability. For the random $4$-regular graph (Fig.~\ref{fig:bicomparisonchart}(g)--(i)), the pair approximation rather than the mean-field approximation again captures the qualitative behavior of the agent-based simulations, with trajectories moving toward an intermediate level of cooperation rather than toward full cooperation. However, the agent-based dynamics quantitatively deviate from the pair approximation dynamics at least in that the dynamics of the agent-based simulation is slower than the pair-approximation dynamics. Because the square lattice also has degree $4$, its pair approximation (Fig.~\ref{fig:bicomparisonchart}(k)) is identical to that for the random $4$-regular graph (Fig.~\ref{fig:bicomparisonchart}(h)). The agreement with the agent-based simulations (Fig.~\ref{fig:bicomparisonchart}($\ell$)) is weaker for the square lattice than for the random $4$-regular graph. This result is expected, since the pair approximation neglects correlations beyond nearest neighbors, which matters more on the square lattice, with its many short loops, than on random regular graphs, which are locally tree-like.

We next examine the dynamics in each region of the $(\omega,\beta)$ phase diagram shown in Fig.~\ref{fig:phasedia}. For each of the four marked parameter values (a)--(d), we initialize the dynamics with $x(0) = 0.5$ and compute $x(t)$ on the same four networks as in Fig.~\ref{fig:bicomparisonchart}.

\begin{figure}[t!]
\centering
\includegraphics[width=140mm]{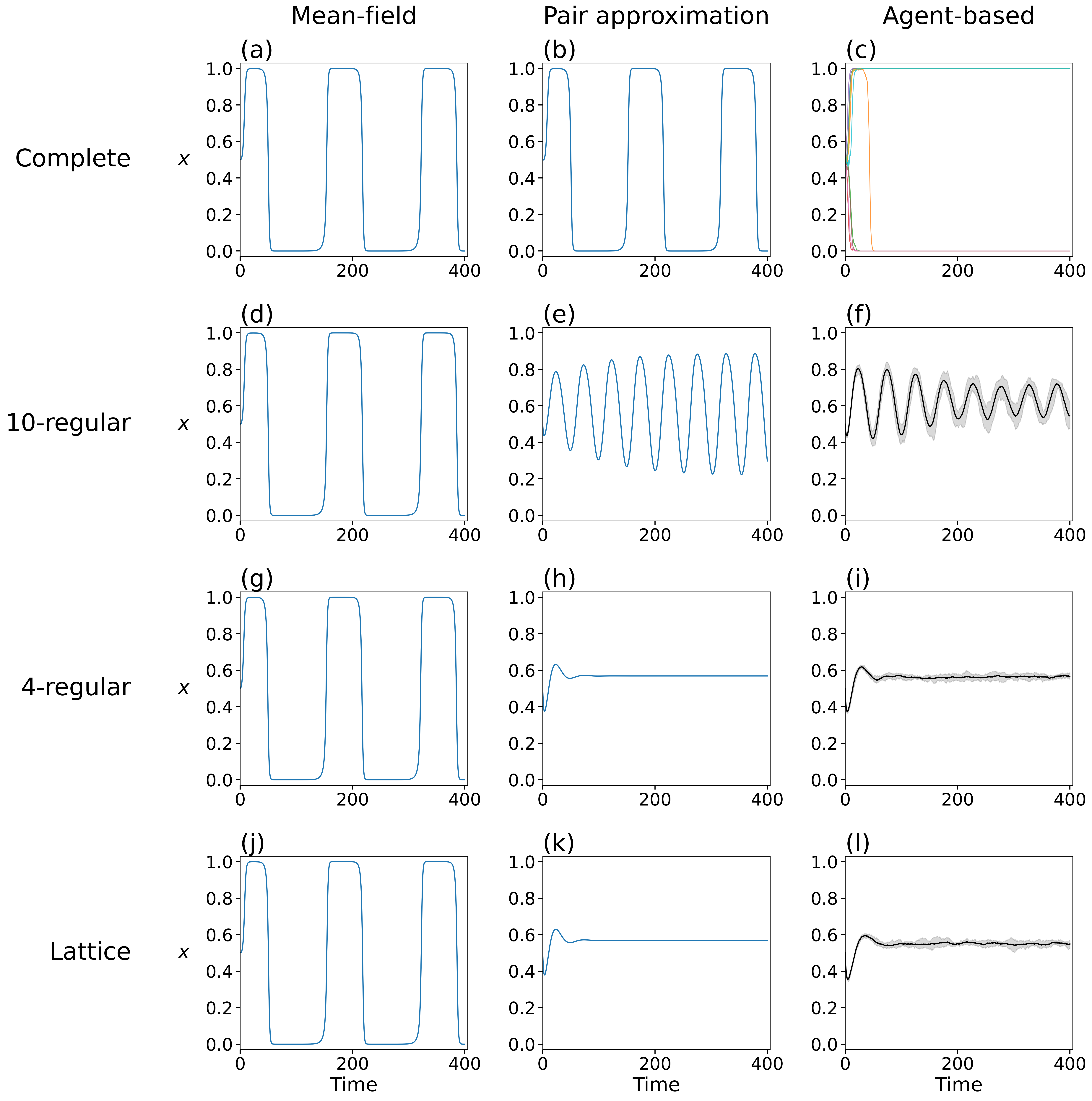}
\caption{Accuracy of the mean-field and pair approximation at point (a) in Fig.~\ref{fig:phasedia}. (a)--(c): Complete graph. (d)--(f): Random $10$-regular graph. (g)--(i) Random $4$-regular graph. (j)--($\ell$): Square lattice. We set $\omega=0.05$, $\beta=5$, and $x(0)=0.5$.  In (c), each line color represents one of the ten runs starting from the same initial condition although many runs nearly overlap on top of each other. In (f), (i), and ($\ell$), the lines show the average over ten runs, and the shaded regions indicate one standard deviation around the average.}
\label{fig:osc(a)comparisonchart}
\end{figure}

Figure~\ref{fig:osc(a)comparisonchart} shows the results for $\omega = 0.05$ and $\beta = 5$, for which the mean-field approximation predicts that the interior equilibrium is a source (point (a) in Fig.~\ref{fig:phasedia}). The mean-field trajectories approach a limit cycle that alternates between sustained periods near full cooperation and sustained periods near full defection (Fig.~\ref{fig:osc(a)comparisonchart}(a), (d), (g), and (j)). For the complete graph, the mean-field (Fig.~\ref{fig:osc(a)comparisonchart}(a)) and pair-approximation (Fig.~\ref{fig:osc(a)comparisonchart}(b)) trajectories agree well with each other. We have confirmed with additional numerical simulations with longer time horizons that the period of oscillation converges as these trajectories undergo more cycles, supporting that the mean-field and pair approximations produce limit cycles. In contrast, the trajectories of the agent-based simulations, shown in different colors for ten different runs in Fig.~\ref{fig:osc(a)comparisonchart}(c), are absorbed into the all-defector or the all-cooperator state at most after one cycle. 
This absorption occurs because the limit cycle passes close to $x=0$ and $x=1$, where demographic noise in a finite population can eliminate one of the two strategies. A similar absorption was reported in stochastic simulations of the original oscillating tragedy-of-the-commons model \cite{Lin19}. Unlike the mean-field and pair approximations, the agent-based simulations do not sustain oscillations for the complete graph.

For the random $10$-regular graph (Fig.~\ref{fig:osc(a)comparisonchart}(d)--(f)), the pair approximation predicts oscillations around an interior state with an amplitude that gradually grows over time. We have confirmed that both the amplitude and period of the oscillation of the pair-approximation trajectory converges after more cycles. The agent-based simulations show damping oscillations with a period similar to that predicted by the pair approximation. Although the damping oscillation behavior is different from eventual limit cycle behavior of the pair-approximation dynamics, the two are similar to each other during the first couple of cycles. The mean-field theory predicts a limit cycle with a substantially larger period and amplitude than the pair-approximation and agent-based-simulation trajectories. These results suggest that the pair approximation is more accurate than the mean-field approximation at describing results of the agent-based simulations. For the random $4$-regular graph and the square lattice (Fig.~\ref{fig:osc(a)comparisonchart}(g)--($\ell$)), the pair approximation predicts damped oscillations that converge to an interior equilibrium with $x \approx 0.57$, closely agreeing with the agent-based simulations, in contrast to the prediction by the mean-field approximation.

\begin{figure}[t!]
\centering
\includegraphics[width=140mm]{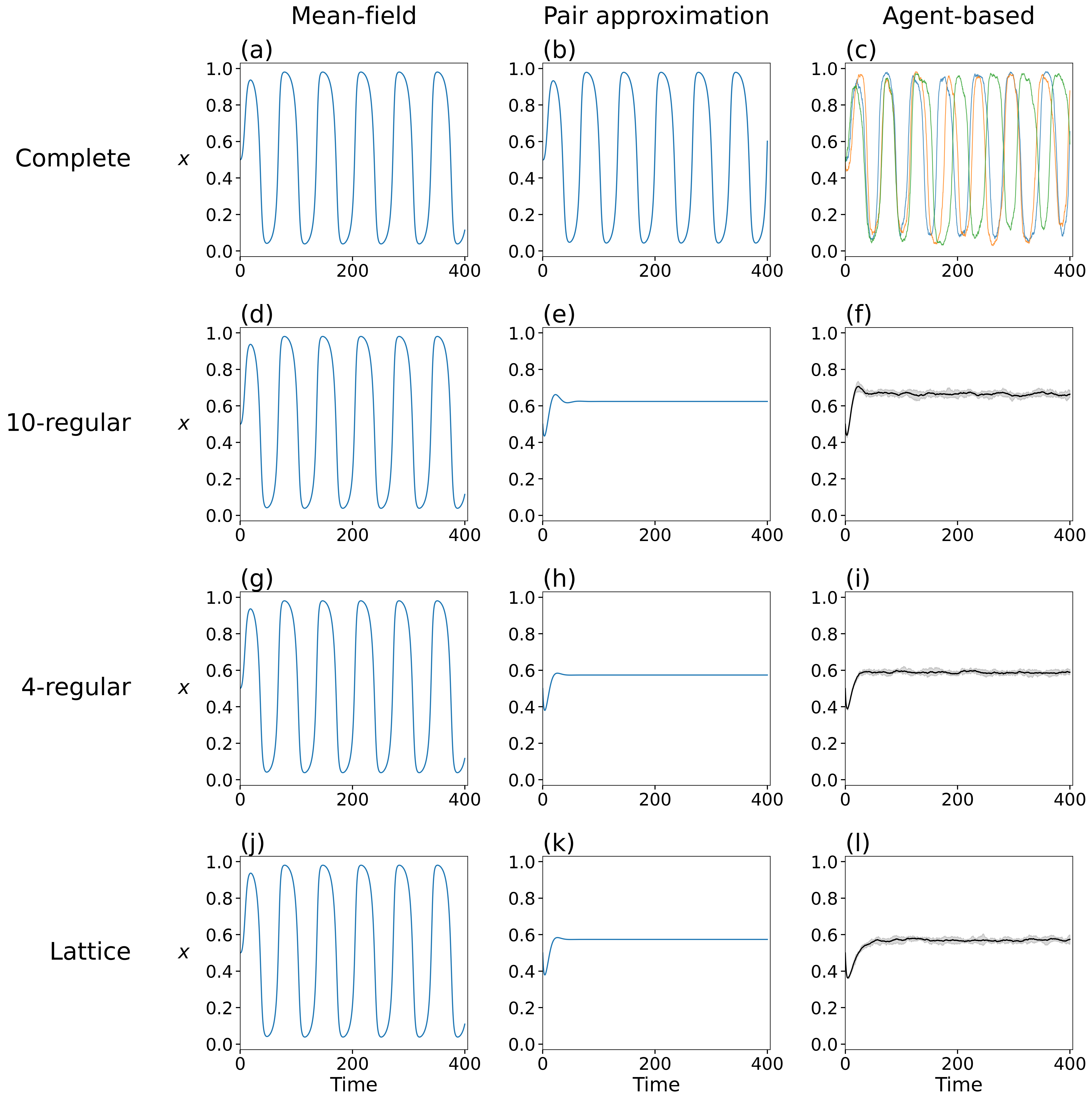}
\caption{Accuracy of the mean-field and pair approximation at point (b) in
Fig.~\ref{fig:phasedia}. (a)--(c): Complete graph. (d)--(f): Random $10$-regular graph. (g)--(i) Random $4$-regular graph. (j)--($\ell$): Square lattice. 
We set $\omega = 0.15$, $\beta=4$, and $x(0)=0.5$. In (c), each line color represents one of the three runs starting from the same initial condition; we plot only three runs because showing trajectories of more runs would make the figure too crowded. In (f), (i), and ($\ell$), the lines show the average over ten runs, and the shaded regions indicate one standard deviation around the average.}
\label{fig:osc(b)comparisonchart}
\end{figure}

Figure~\ref{fig:osc(b)comparisonchart} shows the results for $\omega = 0.15$ and $\beta = 4$, for which the mean-field approximation predicts that the interior equilibrium is a spiral source (point (b) in Fig.~\ref{fig:phasedia}). The mean-field trajectory approaches a limit cycle (Fig.~\ref{fig:osc(b)comparisonchart}(a), (d), (g), and (j)). For the complete graph, the mean-field approximation (Fig.~\ref{fig:osc(b)comparisonchart}(a)) and the pair approximation (Fig.~\ref{fig:osc(b)comparisonchart}(b)) agree well, and both trajectories produce oscillations whose amplitude and period are similar to those of the agent-based simulations (Fig.~\ref{fig:osc(b)comparisonchart}(c)). Unlike in Fig.~\ref{fig:osc(a)comparisonchart}(c), the oscillations from agent-based simulation runs stay far enough from $x = 0$ and $x = 1$ such that the trajectories are not absorbed to $x=0$ or $x=1$ within the simulated time horizon.
For the random $10$-regular graph, the random $4$-regular graph, and the square lattice (Fig.~\ref{fig:osc(b)comparisonchart}(d)--($\ell$)), the pair approximation produces damped oscillations that converge to an interior equilibrium. The agent-based simulations agree with this prediction except that they settle somewhat above the pair-approximation value for the random $10$-regular graph (Fig.~\ref{fig:osc(b)comparisonchart}(f) compared to Fig.~\ref{fig:osc(b)comparisonchart}(e)). In contrast, the trajectory from the mean-field approximation is oscillatory with large amplitudes and therefore far from the trajectories of agent-based simulations for these three networks. Figures~\ref{fig:osc(a)comparisonchart} and \ref{fig:osc(b)comparisonchart} also suggest that sustained or growing oscillations are suppressed on sparser networks.

\begin{figure}[t!]
\centering
\includegraphics[width=140mm]{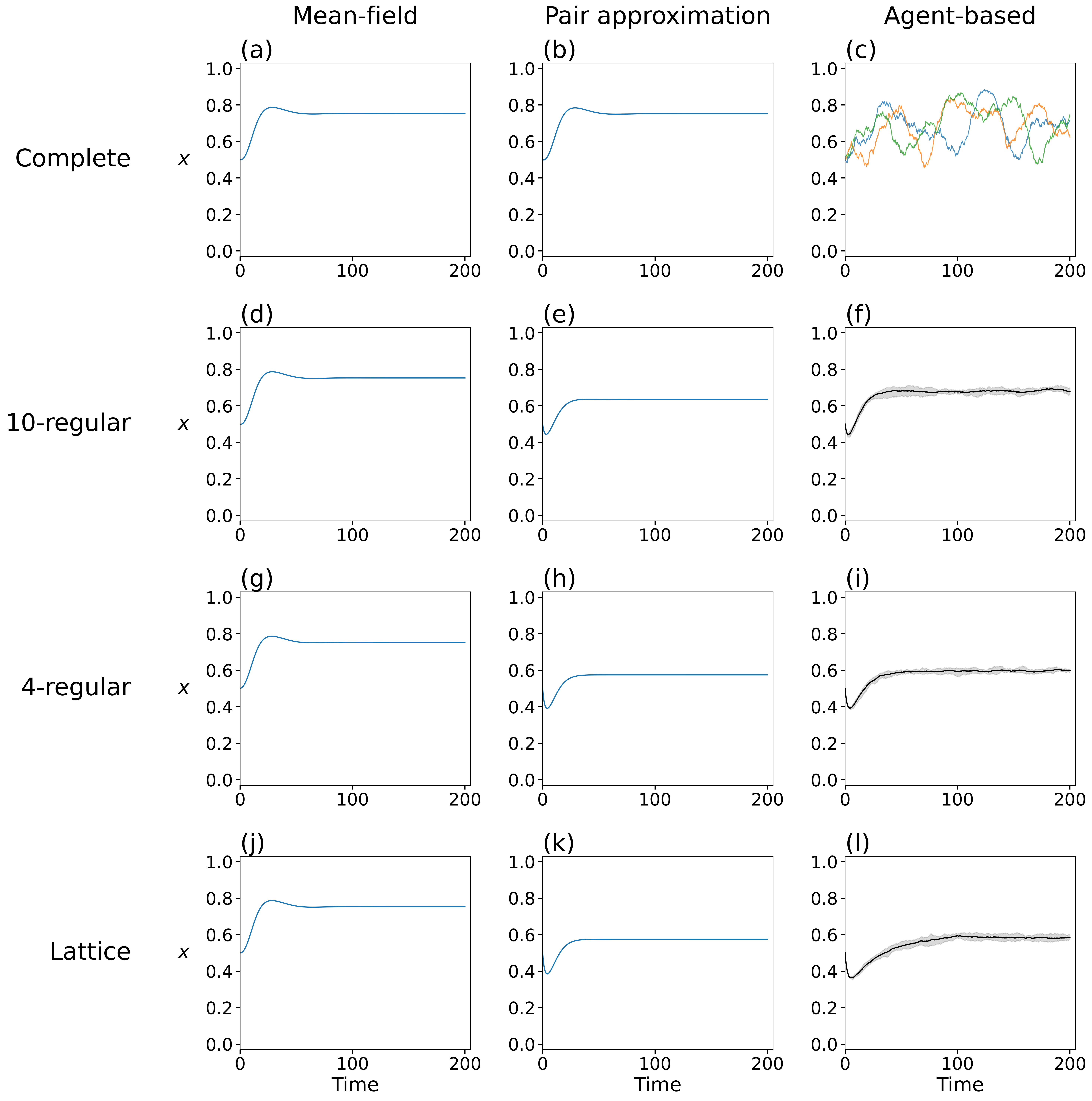}
\caption{Accuracy of the mean-field and pair approximation at point (c) in Fig.~\ref{fig:phasedia}.
(a)--(c): Complete graph. (d)--(f): Random $10$-regular graph. (g)--(i) Random $4$-regular graph. (j)--($\ell$): Square lattice. 
We set $\omega=0.2$, $\beta=2.6$, and $x(0)=0.5$. In (c), each line color represents one of the three runs starting from the same initial condition. In (f), (i), and ($\ell$), the lines show the average over ten runs, and the shaded regions indicate one standard deviation around the average.}
\label{fig:osc(c)comparisonchart}
\end{figure}

Figure~\ref{fig:osc(c)comparisonchart} shows the results for $\omega = 0.2$ and $\beta = 2.6$, for which the mean-field approximation predicts that the interior equilibrium is a spiral sink (point (c) in Fig.~\ref{fig:phasedia}). The mean-field trajectory overshoots slightly and then settles into an interior equilibrium (Fig.~\ref{fig:osc(c)comparisonchart}(a), (d), (g), and (j)). For the complete graph, the pair approximation  (Fig.~\ref{fig:osc(c)comparisonchart}(b)) agrees well with the mean-field approximation (Fig.~\ref{fig:osc(c)comparisonchart}(a)), and both approximations agree reasonably well with the average of agent-based trajectories (Fig.~\ref{fig:osc(c)comparisonchart}(c)). However, the agent-based simulations yield persisting fluctuations around the equilibrium that do not decay over time; see the trajectories of three runs shown in Fig.~\ref{fig:osc(c)comparisonchart}(c). We consider that a larger fluctuation partially owes to the smaller number of nodes, $N=10^3$, for the complete graph, compared to $N = 10^4$ used for the other three networks (Fig.~\ref{fig:osc(c)comparisonchart}(f), (i), and ($\ell$)).
For the other three sparser networks (Fig.~\ref{fig:osc(c)comparisonchart}(d)--($\ell$)), both mean-field and pair approximation qualitatively agree with agent-based simulation results, while the accuracy of approximating the agent-based simulations is higher for the pair than mean-field approximation.

\begin{figure}[t!]
\centering
\includegraphics[width=140mm]{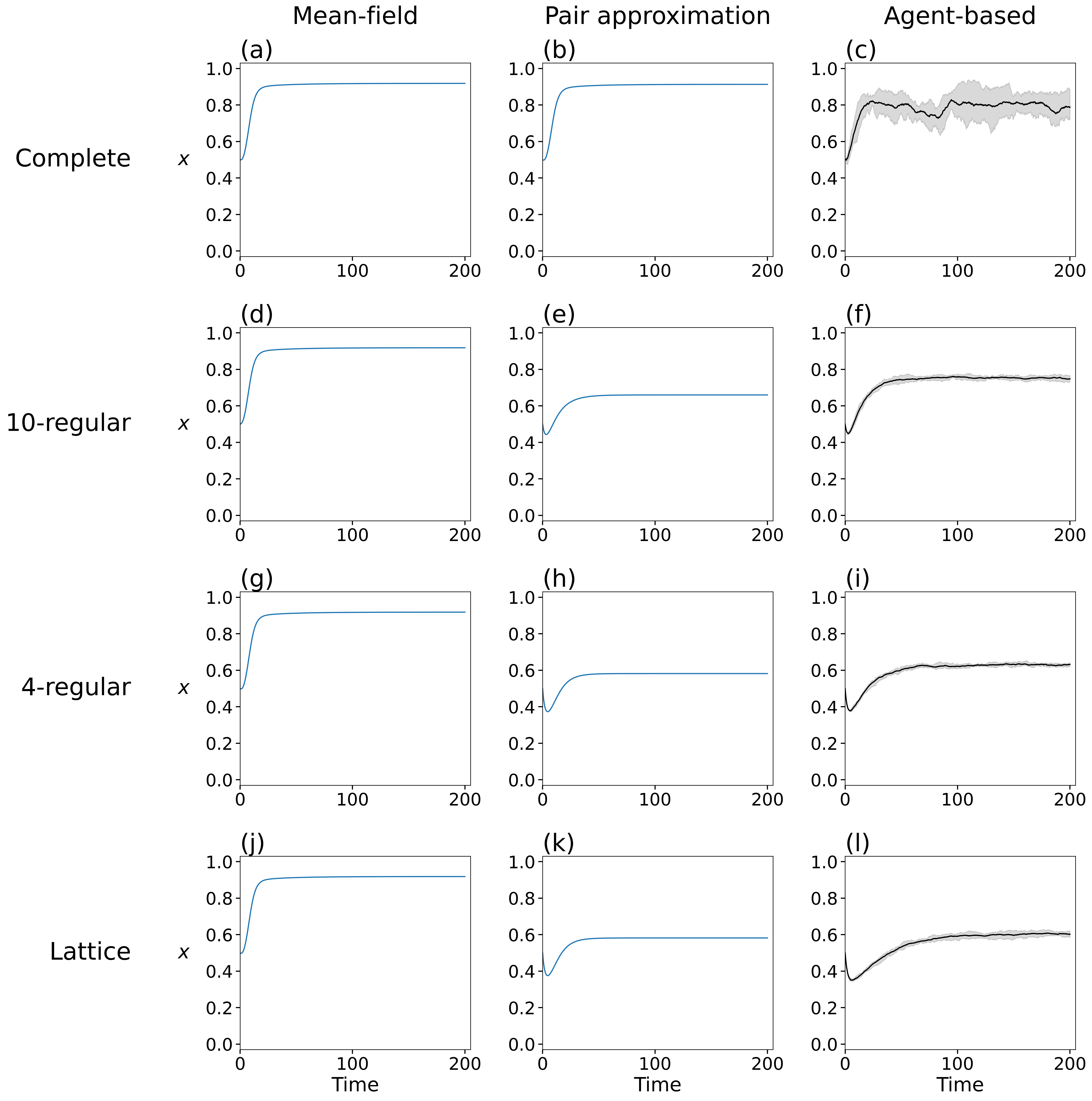}
\caption{Accuracy of the mean-field and pair approximation at point (d) in Fig.~\ref{fig:phasedia}.
(a)--(c): Complete graph. (d)--(f): Random $10$-regular graph. (g)--(i) Random $4$-regular graph. (j)--($\ell$): Square lattice. 
We set $\omega = 0.3$, $\beta=5$, and $x(0)=0.5$. In (c), (f), (i), and ($\ell$), the lines show the average over ten runs, and the shaded regions indicate one standard deviation around the average.}
\label{fig:osc(d)comparisonchart}
\end{figure}

Finally, Fig.~\ref{fig:osc(d)comparisonchart} shows the results for $\omega = 0.3$ and $\beta = 5$, for which the mean-field approximation predicts that the interior equilibrium is a sink (point (d) in Fig.~\ref{fig:phasedia}). The results are similar to those shown in Fig.~\ref{fig:osc(c)comparisonchart}.
In other words, both the mean-field and pair approximations yield qualitatively the same trajectories as the agent-based simulations. In addition, the pair approximation is more accurate than the mean-field approximation at describing trajectories of the agent-based simulations. 

Together, these numerical results show that the mean-field approximation describes the dynamics on the complete graph well but fails on the three sparser networks, where network structure can remove bistability (Fig.~\ref{fig:bicomparisonchart}), suppress oscillations (Figs.~\ref{fig:osc(a)comparisonchart} and \ref{fig:osc(b)comparisonchart}), and lower the equilibrium level of cooperation (Figs.~\ref{fig:osc(c)comparisonchart} and \ref{fig:osc(d)comparisonchart}). The pair approximation captures each of these qualitative changes. Its quantitative accuracy is better than the mean-field approximation 
on the three sparser networks.

\section{Discussion}

We formulated an evolutionary game in which the environment of each interaction is binary and changes in response to the actions of the two players. We analyzed the resulting dynamics using mean-field and pair approximations and agent-based simulations. The mean-field approximation shows that the environmental feedback can produce bistability between full cooperation and full defection, as well as sustained oscillations arising through a Hopf bifurcation. However, the mean-field approximation ignores network structure and therefore predicts the same dynamics on every network. In contrast, the pair approximation tracks the strategies of adjacent nodes jointly with the environmental state of the edge connecting them, so its predictions depend on the node's degree. The pair approximation captures several qualitative changes observed in the agent-based simulations on the sparser networks that the mean-field approximation misses. These changes are the loss of bistability, the suppression of oscillations, and the convergence to a stable interior equilibrium with a lower level of cooperation. The quantitative accuracy of the pair approximation is lower on the square lattice than on the random regular graphs because the square lattice has many short loops. Together, these results show how strategy-dependent environmental change and population structure jointly shape evolutionary dynamics.

Previous studies of replicator dynamics with environmental feedback have reported heteroclinic cycles on the boundary of the phase space, both in well-mixed populations \cite{Weitz16, Kawano19, Ma24} and on regular graphs \cite{Wang25}, and have found that limit cycles cannot occur in the original well-mixed model \cite{Gong22}. In these models, the fully replete and fully depleted environments are invariant, so the corners of the phase space are saddles connected along the boundary. Changing the environmental dynamics can replace the heteroclinic cycle with a stable limit cycle, for example when the environment has intrinsic growth, decay, or a tipping point \cite{Tilman20}. Our model also changes the environmental dynamics from those models. Specifically, each edge switches between binary environmental states with probabilities that depend on the strategies at its endpoints (see Fig.~\ref{fig:schematic}), so the environment never becomes fully replete or fully depleted. Consistent with this observation, the mean-field dynamics of our model converge to a stable limit cycle that passes close to the boundary but has a constant period and amplitude, rather than converge to a heteroclinic cycle. Under the pair approximation, decreasing the node's degree, $k$, can shrink the limit cycle away from the boundary on random 10-regular networks (compare Fig.~\ref{fig:osc(a)comparisonchart}(e) with Fig.~\ref{fig:osc(a)comparisonchart}(b)) and replaces it with a stable interior equilibrium on random 4-regular networks (Fig.~\ref{fig:osc(a)comparisonchart}(h)) and the square lattice (Fig.~\ref{fig:osc(a)comparisonchart}(k)).

Beyond well-mixed populations, square-lattice simulations have investigated how interaction history changes the environmental state of individual edges \cite{He23,Zhu23}. These models modify the temptation to defect through variables representing strategy persistence or the history of cooperation, and show that environmental variation can promote cooperation \cite{He23} or either promote or hinder it depending on the parameters \cite{Zhu23}. Our study complements these numerical investigations by considering binary edge environments with negative feedback and developing a pair approximation that captures qualitative changes across network structures, including the loss of bistability and suppression of oscillations. A natural extension of our model in this direction is to let environmental switching depend explicitly on the duration or recent history of the interacting players' strategies.

Our stability analysis of the all-defector and all-cooperator equilibria uses an approximate rare-invader calculation. At each boundary, some pair types are absent, so their environmental fractions are treated as auxiliary quantities and estimated from steady states of the limiting equations. This ansatz provides practical invasion conditions. However, a fuller mathematical treatment either justifying this limiting procedure or building a more rigorous alternative method warrants future work. For example, one may be able to establish when the conditional environmental fractions, such as $M_{CC}^{(0)}$, $M_{CD}^{(0)}$, and $M_{DD}^{(0)}$, converge near each boundary and develop bounds on the resulting mutant growth rate to prove whether sufficiently small invasions decay or grow.

Our model is closely related to the game-transition framework of Su et al.~\cite{Su19GT}, in which the game played on an edge changes according to its current state and the strategies of its endpoints and analytical results are derived under weak selection limit. Their principal examples involve transitions between donation games that favor defection individually, with mutual cooperation leading to a more valuable game and thereby improving the prospects for cooperation. In our model, the two environmental states instead favor opposite strategies. Specifically, cooperation drives an edge toward a rich environment that favors defection, whereas defection drives it toward a depleted environment that favors cooperation. We examined how this negative feedback and network structure jointly shape bistability, coexistence, and oscillations at finite selection strength, complementing their weak-selection analysis of fixation. A potentially promising future study is to adapt their pair approximation and diffusion approach to calculate fixation probabilities under weak selection in our setting. Such an analysis could determine how the relative frequency of environmental updating and node's degree affect how likely a single cooperator takes over a defector population, complementing the equilibrium and stability results obtained in this study.

\section*{Acknowledgments}

%\kb{Ask Feng for his funding acknowledgements}
N.M. acknowledges support by the NSF under Grant No.\,DMS-2204936, JSPS KAKENHI under Grants No.\,23H03414, No.\,24K14840, and No.\,24K030130, and by Japan Science and Technology Agency (JST) under Grant No.\,JPMJMS2021.
During the preparation of this manuscript, the authors used ChatGPT-6.1 Sol for manuscript editing, ChatGPT Pro 5.5 for assisting the theoretical calculation in sections~\ref{sec:equilibria} and \ref{sec:pa}, and Claude Opus 5.5 for code editing.

\begin{appendices}

\section{Jacobian of the mean-field dynamics}\label{app:MF1}

In this section, we compute the Jacobian of the four-dimensional mean-field system given by Eqs.~\eqref{eq:dxrho}, \eqref{eq:dpcc}, \eqref{eq:dpcd}, and \eqref{eq:dpdd}. The formulas below apply for $0<x<1$, where the payoff functions in Eqs.~\eqref{eq:FCinf} and \eqref{eq:FDinf} are differentiable.

For compactness, write
$u=P_{CC}^{(0)}$, $v=P_{CD}^{(0)}$, $z=P_{DD}^{(0)}$, $\alpha=1-\omega$,
$\Delta=F_C-F_D$, $\rho=\frac{1}{1+\exp(-\beta\Delta)}$, 
$\bar\rho=1-\rho$, $h=2\rho-1$, and $\chi=\beta\rho\bar\rho$.
The partial derivatives of the payoff difference, $\Delta$, with respect to $x$ while we held $u$, $v$, and $z$ fixed is given by
\begin{align}
\Delta_x \equiv \frac{\partial \Delta}{\partial x} ={}&R_1-S_1-T_1+P_1
-\frac{(R_0-R_1)u}{x^2}
-\frac{(S_0-S_1)v}{2x^2}\notag\\
&-\frac{(T_0-T_1)v}{2(1-x)^2}
-\frac{(P_0-P_1)z}{(1-x)^2}.
\label{eq:Deltax}
\end{align}
Similarly, we obtain
\begin{align}
\Delta_u={}&\frac{R_0-R_1}{x},\label{eq:Deltau}\\
\Delta_v={}&\frac{S_0-S_1}{2x}-\frac{T_0-T_1}{2(1-x)},\label{eq:Deltav}\\
\Delta_z={}&-\frac{P_0-P_1}{1-x}.\label{eq:Deltaz}
\end{align}
In addition, we find $\partial\rho/\partial\xi=\chi\Delta_\xi$ for $\xi\in\{x,u,v,z\}$. We also define
\begin{align}
\Psi_1&=xv+2(1-x)u,\label{eq:Psi1}\\
\Psi_2&=x(2z-v)-(1-x)(2u-v),\label{eq:Psi2}\\
\Psi_3&=(1-x)v+2xz.\label{eq:Psi3}
\end{align}

With the variable order $(x,u,v,z)$, the Jacobian is
\begin{equation}\label{MFGenJac}
J=
\begin{pmatrix}
J_{11}&J_{12}&J_{13}&J_{14}\\
J_{21}&J_{22}&J_{23}&J_{24}\\
J_{31}&J_{32}&J_{33}&J_{34}\\
J_{41}&J_{42}&J_{43}&J_{44}
\end{pmatrix},
\end{equation}
where
\begin{align}
J_{11}&=\alpha\left[(1-2x)h+2x(1-x)\chi\Delta_x\right],\label{eq:J11}\\
J_{12}&=2\alpha x(1-x)\chi\Delta_u,\label{eq:J12}\\
J_{13}&=2\alpha x(1-x)\chi\Delta_v,\label{eq:J13}\\
J_{14}&=2\alpha x(1-x)\chi\Delta_z,\label{eq:J14}\\
J_{21}&=4\omega q_0x+\alpha\left[v\rho+2u\bar\rho+\chi\Psi_1\Delta_x\right],\label{eq:J21}\\
J_{22}&=-2\omega(p_0+q_0)
+\alpha\left[-2(1-x)\bar\rho+\chi\Psi_1\Delta_u\right],\label{eq:J22}\\
J_{23}&=\alpha\left[x\rho+\chi\Psi_1\Delta_v\right],\label{eq:J23}\\
J_{24}&=\alpha\chi\Psi_1\Delta_z,\label{eq:J24}\\
J_{31}&=4\omega q_1(1-2x)
+\alpha\left[(2z-v)\rho-(2u-v)\bar\rho+\chi\Psi_2\Delta_x\right],\label{eq:J31}\\
J_{32}&=\alpha\left[2(1-x)\bar\rho+\chi\Psi_2\Delta_u\right],\label{eq:J32}\\
J_{33}&=-2\omega(p_1+q_1)
+\alpha\left[-x\rho-(1-x)\bar\rho+\chi\Psi_2\Delta_v\right],\label{eq:J33}\\
J_{34}&=\alpha\left[2x\rho+\chi\Psi_2\Delta_z\right],\label{eq:J34}\\
J_{41}&=-4\omega q_2(1-x)
+\alpha\left[-v\bar\rho-2z\rho-\chi\Psi_3\Delta_x\right],\label{eq:J41}\\
J_{42}&=-\alpha\chi\Psi_3\Delta_u,\label{eq:J42}\\
J_{43}&=\alpha\left[(1-x)\bar\rho-\chi\Psi_3\Delta_v\right],\label{eq:J43}\\
J_{44}&=-2\omega(p_2+q_2)
+\alpha\left[-2x\rho-\chi\Psi_3\Delta_z\right].\label{eq:J44}
\end{align}
One should use Eqs.~\eqref{MFGenJac}--\eqref{eq:J44} only for interior equilibria. Stability calculations for the boundary equilibria (i.e., $x=0$ or $x=1$) 
in section~\ref{sec:equilibria} use the nonsingular variables $(x,c_1,c_2,c_3)$ and Eqs.~\eqref{eq:regx}--\eqref{eq:regc3}, rather than substituting $x=0$ or $x=1$ directly into this Jacobian.

\end{appendices}

\bibliographystyle{unsrt}
\bibliography{Betz_Research_Refs}
%%%%%%%%%%%%%%%%%%%%%%%Appendix A%%%%%%%%%%%%%%%%%%%%%

\end{document}